\documentclass[a4paper, 11pt]{article}
\usepackage[pdftex, a4paper]{geometry}
\usepackage{graphicx}				% Use pdf, png, jpg, or eps¡ì with pdflatex; use eps in DVI mode
\usepackage{sansmath}
\usepackage{amsmath}
\usepackage{amsfonts}
\usepackage{dsfont}
\usepackage{mathrsfs}
\usepackage{amssymb}
\usepackage{bbm}
\usepackage{epsfig}
\usepackage[english]{babel}
\usepackage[all]{xy}
\usepackage{subfigure}
\usepackage{color}
\usepackage{cases}

\newcommand\dc[1]{\discretionary{#1}{}{}}

\newcommand*\mcapinn[2]{\vcenter{\hbox{$\mathsurround=0pt
  \ifx\displaystyle#1\textstyle\else#1\fi\bigcap$}}}

\newcommand*\mcup{\mathbin{\mathpalette\mcupinn\relax}}
\newcommand*\mcupinn[2]{\vcenter{\hbox{$\mathsurround=0pt
  \ifx\displaystyle#1\textstyle\else#1\fi\bigcup$}}}

\DeclareFontFamily{OT1}{pzc}{}
\DeclareFontShape{OT1}{pzc}{m}{it}{<-> s * [1.200] pzcmi7t}{}
\DeclareMathAlphabet{\mathpzc}{OT1}{pzc}{m}{it}

\newtheorem{theorem}{Theorem}
\newtheorem{definition}{Definition}

\newtheorem{lemma}{Lemma}

\newsavebox{\mycases}

\begin{document}

\title{\bf Feedback Capacity over Networks}

\author{Bo Li and  Guodong Shi\thanks{The two authors contributed equally to this work. The authors are with the Research School of Engineering, The Australian National University, ACT 0200,
Canberra, Australia. Email:
libo@amss.ac.cn, guodong.shi@anu.edu.au.}}

\date{}
\maketitle
\begin{abstract}
In this paper, we investigate the fundamental limitations of feedback mechanism in dealing with uncertainties for network systems. The study of maximum capability of feedback control was pioneered in Xie and Guo (2000) for  scalar systems with nonparametric nonlinear uncertainty. In a network setting,  nodes with unknown and nonlinear   dynamics  are interconnected through a directed interaction graph.  Nodes can design  feedback controls  based on all available information, where the objective is to stabilize the network state.  Using information structure and decision pattern as  criteria, we specify three  categories  of network feedback laws, namely the global-knowledge/\dc{}global-decision, network-flow/local-decision, and local-flow/local-decision feedback. We establish
a series of network capacity characterizations   for these three basic types of network control laws. First of all,  we prove that for  global-knowledge/\dc{}global-decision  and network-flow/local-decision control where nodes know the information flow across the entire network, there exists a critical number  $
\big(3/2+\sqrt{2}\big)/\|A_{\mathrm{G}}\|_\infty
$,
where $3/2+\sqrt{2}$ is as known as the Xie-Guo constant and $A_{\mathrm{G}}$ is the network adjacency matrix, defining exactly how much uncertainty in the node dynamics can be overcome by feedback.  Interestingly enough, the same feedback capacity  can be achieved  under max-consensus enhanced local flows where nodes only observe information flows from neighbors as well as extreme (max and min) states in the network. Next, for local-flow/local-decision control, we prove that there exists a structure-determined value being a lower bound of the network feedback capacity. These results reveal the important  connection between  network structure and fundamental capabilities   of in-network  feedback control.
\end{abstract}


\section{Introduction}
\subsection{Background}
Lying  at  the heart of practicing  and understanding  control systems has been  the feedback mechanism. Today it is recognized that the first systematic study of feedback control was made by  J. C. Maxwell  in 1868  on   pendulum    governors \cite{Maxwell-1968}. Invented  by J. Watt   to control his steam engine, the so-called fly-ball  governor senses engine speed via the spinning angle of two weighted balls,  and in the mean time adjusts the steam valve through levers connected to the balls \cite{Franklin-Book-2015}.  The basic  idea of feedback has since been clear from this historical  example: the behaviour of a dynamical system can be regulated by feeding  the outputs of the system back to its inputs, and particularly, via feedback  unknown disturbances can be rejected to a desired level at the output end. How to design and optimize feedback controllers that can maximally  reduce the effects of internal or external  uncertainty becomes a central theme in the field   of automatic control \cite{Basar-Book-1991}.

The influence of external uncertainty such as  disturbances and sensor  noises  can be well and conveniently  understood by classical frequency-based methods \cite{Franklin-Book-2015}. Treatments to internal and structural uncertainties  that are  ubiquitous in real-world plants are however far more challenging. There are two parallel but related major  research paths along which  celebrated results have been developed for discrete-time or continuous-time, linear or nonlinear, and autonomous or time-varying systems.  Robust control synthesis \cite{Zames-1966,Doyle-1982,Khammash-1991,Zhou-Book-1996,Matt-Book-1999} characterizes uncertainty within   a prescribed (often compact) set around  the true plant, and controllers are designed often to optimize   certain performance metrics induced by the uncertainty neighborhood, e.g., maximizing performance for worst-case scenarios. Adaptive control methodology  \cite{Astrom-Automatica-1973,Mareels-Automatica-1992,Mareels-Automatica-1998,Byrnes,Astrom-Book-1995,Krstic-Book-1995} utilizes online estimation techniques from  the input-output signals, where controllers are adjusted in real time from the estimation outcomes.

The study of feedback control  has been pushed forward to a new network era in the past decade, inspired by the emergence of a variety of dynamical systems  of  complex networks. The need of carrying out control and sensing over communication channels  has led to the introduction of information theory to the study of control systems.  In-depth results have been established for the necessary data rate between the sensor  and actuator for stabilizing a plant \cite{Wong-TAC-1999,Nair-SIAM-2004,Liberzon-TAC-2005,Nair-PIEEE-2007,Nair-TAC-2013}, and for the performance of control and estimation over  lossy or noisy channels  \cite{Schenato-PIEEE-2007,Dahleh-TAC-2008}. Moreover,  the notion of distributed control \cite{Jadbabaie-TAC-2003}  sparkled a tremendous amount of work aiming at robust and  scalable solutions for a large number of interconnected nodes to achieve collective goals ranging from consensus and formation to optimization and  computation \cite{saber2004,Bullo-TAC-2007,Nedic-TAC-2010,Mou-TAC-2015,Shi-TAC-LAE}. Multi-agent control has evolved to a discipline in its own right \cite{Magnus-Book-2010}, being generalized even to control of quantum networks \cite{Shi-TAC-Quantum}. Of particular interest there is also the study of network  controllability    \cite{Magnus-SIAM-2009,Barabasi-Nature-2011,Olshevsky-TCNS-2014}, focusing on how interaction  structures influence network  controllability when measurement and control take place at a few selected nodes.

\subsection{Motivation}
Besides the tremendous  success of in-network control design  \cite{Magnus-Book-2010}, it is equally important to understand the limitations of feedback mechanism over network dynamics facing uncertainty.  More specifically, a clear characterization to  the capacity of feedback mechanism over a network in dealing with uncertainty, for centralized and distributed controllers, respectively, will help us understand the boundaries of controlling complex networks from a theoretical perspective.

In the seminal work \cite{xie-guo2000}, Xie and Guo established foundational  results on  the capability of feedback mechanism with nonparametric nonlinear uncertainty for the following discrete-time model
\begin{align*}
\mathbf{y}(t+1)=f(\mathbf{y}(t))+\mathbf{u}(t)+ \mathbf{w}(t), \ t=0,1,\dots
\end{align*}
where the $\mathbf{y}(t)$, $\mathbf{u}(t)$, and $\mathbf{w}(t)$ are real numbers representing output, control, and disturbance, respectively.  It was shown in \cite{xie-guo2000} that with completely unknown plant model $f(\cdot):\mathbb{R}\to \mathbb{R}$ and bounded but unknown disturbance signal $\mathbf{w}(t)$, a necessary and sufficient condition for the existence of stabilizing feedback control of the above system is that a type of Lipschitz norm of $f(\cdot)$ must be strictly smaller than $3/2+\sqrt{2}$.

This number,  now referred to as the {\it Xie-Guo constant} in the literature,  points to the ultimate  limitations of {\it all} feedback laws. Generalizations have been made  for a few types of parametric models for which the corresponding feedback capabilities can be characterized  \cite{Li-Guo-Automatica-2010,Li-Guo-TAC-2011,Huang-Guo-Automatica-2012,Sokolov-Automatica-2016}. Naturally we wonder (i) Would such a feedback capacity critical value exist for a network system? (ii) If it indeed exists, how would it depend on the network structure? (iii) How feedback capacity would differ between  centralized and distributed controllers? Answers to these questions will add to important  understandings for control of networked  systems and for  feedback mechanism itself as well.
\subsection{Main Results}
We consider a network setting of  the  nonparametric  uncertainty model in  \cite{xie-guo2000}, where nodes with unknown nonlinear    self-dynamics are interconnected through a directed interaction graph. For the ease of presentation the dynamics of the nodes are assumed to be identical, corresponding to homogenous networks. The interaction graph defines neighbor relations among the nodes, where measurement and control take place. Nodes can design any feedback controller using the information they have, and the objective is to stabilize the entire network, i.e., every node state in the network.

Three basic categories  of feedback laws over such networks are carefully specified.  In global-knowledge/\dc{}global-decision feedback, every node knows  network structure (interaction graph) and   network information flow, and  nodes can coordinate to make control decisions; in network-flow/local-decision feedback, each node only knows the network information flow and carries out decision individually; in local-flow/local-decision feedback, nodes only know information flow of neighbors and then make their own control decisions. Note that various  existing distributed controllers and algorithms can be naturally  put into  one of the three categories.  A series of network feedback capacity results has been established:
\begin{itemize}
\item[(i)] For  global-knowledge/\dc{}global-decision  and network-flow/local-decision control,   the generic network  feedback capacity  is fully captured by a critical value   $$
\big(3/2+\sqrt{2}\big)/\|A_{\mathrm{G}}\|_\infty
$$
where $A_{\mathrm{G}}$ is the network adjacency matrix.

\item[(ii)] For local-flow/local-decision control, there exists a structure-determined value being an lower bound of the network feedback capacity.

\item[(iii)] Network flow can be replaced by max-consensus enhanced local flows, where nodes  only observe information flows from their neighbors as well as  network extreme (max and min) states  via max-consensus, and then  the same feedback capacity  can be reached.
\end{itemize}
Additionally, for strongly connected graphs, we manage to establish a universal impossibility theorem on the existence of stabilizing feedback laws.
\subsection{Paper Organization}
The remainder of the paper is organized as follows. Section \ref{sec:model} introduces the network model and defines the problem of interest. Section \ref{sec:results} presents the main results, followed by Section \ref{sec:controllers} presenting the network stabilizing  controllers. Section \ref{sec:proofs} provides the proofs of all the statements. Finally Section \ref{sec:conclusions} concludes the paper by a few remarks  pointing out a few interesting future directions.

\medskip

\noindent {\em Notation}: The set of real numbers is denoted by $\mathbb{R}$, and the set of integers  is denoted by $\mathbb{Z}$. A sequence $a_0,a_1,\dots$ is abbreviated as $\langle a_t \rangle_{t\geq 0}$. For any real number $a$, $(a)^{+}$ is defined as $(a)^{+}=\max\{a,0\}$.  For convenience we use ${\rm dist}(X,Y)$ to denote the distance between two sets $X$ and $Y$ in $\mathbb{R}$ by ${\rm dist}(X,Y)=\inf_{x\in X, y\in Y} |x-y|$, and simply  ${\rm dist}(a,Y):=\inf_{y\in Y} |a-y|$, ${\rm dist}(a,b)=|a-b|$ for real numbers $a$ and $b$.
\section{The Model}\label{sec:model}

\subsection{Network Dynamics with Uncertainty}
Consider a network with $n$ nodes indexed in the set $\mathrm{V}=\{1,..., n\}$. The network interconnection structure is represented  by a  directed graph $\mathrm{G} = (\mathrm{V}, \mathrm{E})$, where $\mathrm{E}$ is the arc set.   Each arc $(i,j)$ in the set $\mathrm{E}$ is an ordered pair of two nodes $i,j\in\mathrm{V}$, and link $(i,i)$  is allowed at each node $i$ defining a self-arc. The neighbors of node $i$, that node $i$ can be influenced by,  is defined as nodes in the set $\mathrm{N}_i:=\{j:(j,i)\in\mathrm{E}\}$. Let $a_{ij}\in\mathbb{R}$ be a real number representing the weight of the directed arc $(j,i)$ for $i,j\in\mathrm{V}$. The arc weights $a_{ij}$ comply with the network structure in the sense that $a_{ij}\neq 0$ if and only if $(j,i)\in \mathrm{E}$. Let $A_{\mathrm{G}}$ be the adjacency matrix  of the graph $\mathrm{G}$ with $[A_{\mathrm{G}}]_{ij}=a_{ij}$.

Time is slotted at $t=0,1,2,\dots$. Each node $i$ holds a state $\mathbf{s}_i(t)=\big(\mathbf{x}_i(t),\mathbf{z}_i(t)\big)^\top\in \mathbb{R}^2$ at time $t$. The network dynamics are described by
\begin{equation}\label{eqn:evol}
\begin{aligned}
&\mathbf{z}_i(t+1)=f(\mathbf{x}_i(t))+\mathbf{e}_i(t)\\
&\mathbf{x}_i(t+1) = \sum_{j\in \mathrm{N}_i} a_{ij} \mathbf{z}_j(t+1)+ \mathbf{u}_i(t) + \mathbf{w}_i(t),\\
\end{aligned}
\end{equation}
for $i\in\mathrm{V}$ and $t=0,1,2,\dots$, where  ${f}$ is a function mapping  from $\mathbb{R}$ to $\mathbb{R}$,  $\mathbf{u}_i(t)\in\mathbb{R}$ is the control input,  and $\mathbf{e}_i(t), \mathbf{w}_i(t)\in\mathbb{R}$ are disturbances and noises. The system (\ref{eqn:evol}) describes the following node interactions: $\mathbf{x}_i(t)$ is the internal state of node $i$ at time $t$, based on which an external state $\mathbf{z}_i(t+1)$ is generated at that node; at time $t+1$, the external states $\mathbf{z}_i(t+1)$ are exchanged over the interaction  graph $\mathrm{G}$, defining the update of  the internal states $\mathbf{x}_i(t+1)$. In this way,  $\big(\mathbf{x}_i(t),\mathbf{z}_i(t+1)\big)$ is an input-output pair at node $i$ for time $t$. We impose the following standing assumptions.

\medskip

\noindent {\bf Assumption 1}. ({\em Dynamics Uncertainty}) The function $f$ is unknown, and  the arc weight $a_{ij}$ is known to the node $i$.

\medskip

\noindent{\bf Assumption 2}. ({\em  Disturbance Boundedness})  The process disturbances $\mathbf{e}_i(t)$ and  $\mathbf{w}_i(t)$ are unknown but bounded, i.e., there exist $e_\ast, w_\ast>0$ such that $$
\big|\mathbf{e}_i(t)\big|\leq e_\ast, \quad \big|\mathbf{w}_i(t)\big|\leq w_\ast
$$ for all $t$ and for all $i\in\mathrm{V}$. Furthermore, the bounds $e_\ast$ and $w_\ast$ are unknown.

\medskip

\begin{figure}
\centering
\includegraphics[width=5.4in]{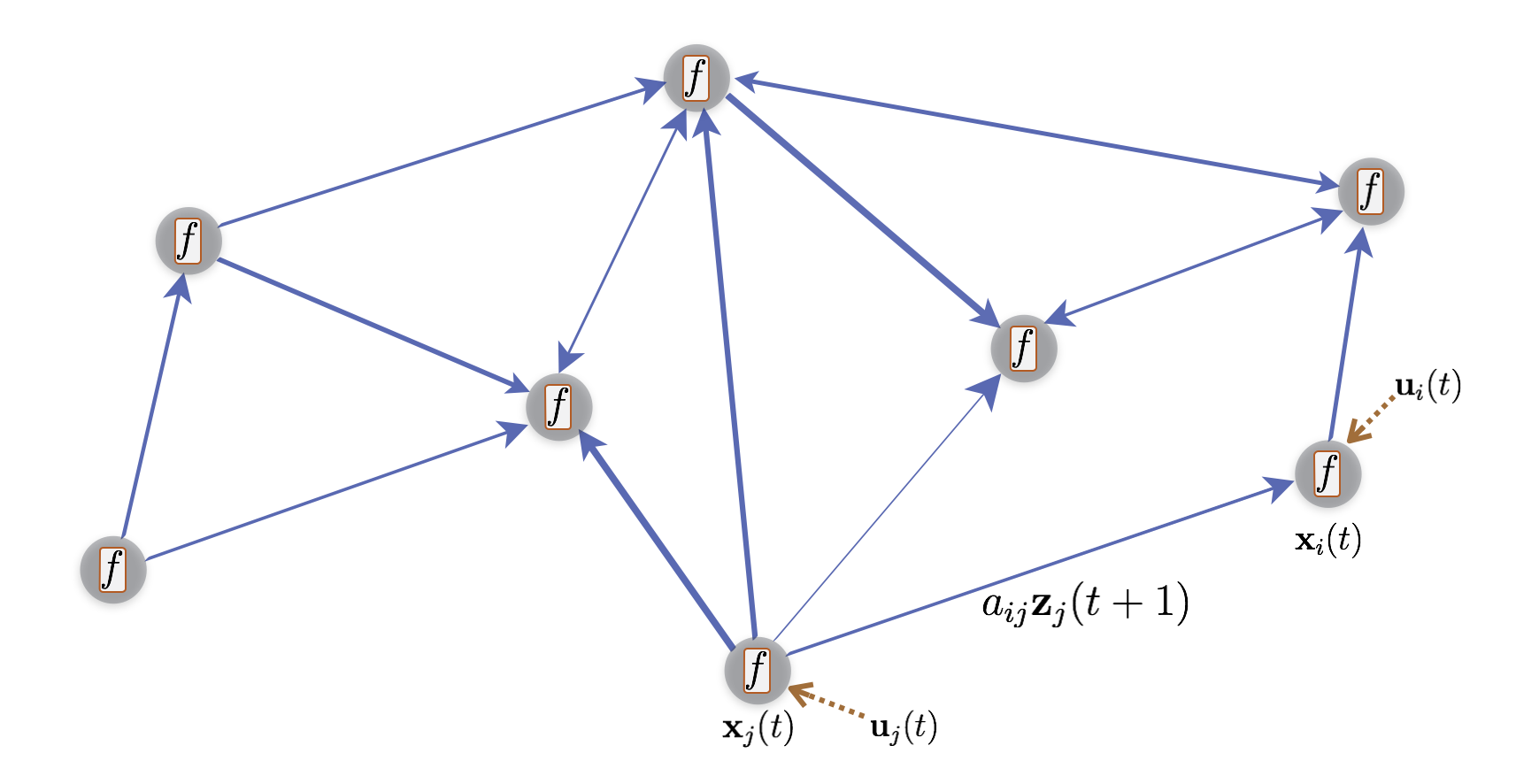}
\caption{A graphical diagram of the  considered network model: (i) Interaction structure  forms a directed graph where nodes are influenced by their in-neighbors and influence their out-neighbors; (ii) Node interaction rules are governed by completely unknown nonlinear dynamics and link width indicates the weight (strength) of interactions; (iii) Control inputs are applied to individual nodes subject to unknown disturbances.  }
\label{fig:diagram}
\end{figure}

The Assumptions 1--2 are quite natural and general, which  are adopted throughout the remainder of the paper without specific further mention. An illustration of this dynamical network model can be seen in Fig. \ref{fig:diagram}. The dynamics of the internal node states $\mathbf{x}_i(t)$ can be written in a compact form as
\begin{equation}\label{eqn:evol_internal}
\begin{aligned}
&\mathbf{x}_i(t+1) = \sum_{j\in \mathrm{N}_i} a_{ij} f(\mathbf{x}_j(t))+ \mathbf{u}_i(t) + \mathbf{d}_i(t),\\
\end{aligned}
\end{equation}
where $\mathbf{d}_i(t)=\sum_{j\in \mathrm{N}_i} a_{ij}\mathbf{e}_j(t) + \mathbf{w}_i(t)$. 

\subsection{Feedback Laws over Networks}
We now   classify all possible  network feedback  control laws into categories determined by {\it information patterns} and {\it decision structures}. Such a classification is not straightforward at all bearing the following questions in mind:
\begin{itemize}
\item[(i)] ({\em Knowledge}) How much would nodes know about the network itself, e.g., number of nodes $n$, network connectivity, or even the network topology $\mathrm{G}$?

 \item[(ii)] ({\em Flows})   How much would  nodes know about the network information flows, e.g.,  availability of $\mathbf{x}_i(t)$, $\mathbf{z}_i(t)$, and $\mathbf{u}_i(t)$ for a neighbor, or a neighbors' neighbor of the node $i$?

    \item[(iii)]  ({\em Decisions})  To what level  nodes could cooperate in determining the control actions, e.g., can a node $i$ tell a neighbor $j$ to stand by with $\mathbf{u}_j(t)=0$ at time $t$ to implement its own control input $\mathbf{u}_i(t)$?
 \end{itemize}
Different answers to these questions will lead to drastically different scopes of network control rules. In this paper, we focus on a few fundamental forms of network feedback laws that from a theoretical perspective  represent a variety of network control and computation results in the literature.

Denote $\mathbf{S}(t)=(\mathbf{s}_1(t)^\top \dots \ \mathbf{s}_n(t)^\top)^\top$  and $\mathbf{U}(t)=(\mathbf{u}_1(t)\ \dots \ \mathbf{u}_n(t))^\top$ for $t=0,1,\dots$. Here without loss of generality we assume $\mathbf{z}_i(0)=0$ for all $i$.  The following  definition specifies network and local flows.

\begin{definition} The network flow vector up to time $t$ is defined as
\begin{align*}
 \Theta(t):=\Big(\mathbf{S}(0),\dots, \mathbf{S}(t); \mathbf{U}(0),\dots,\mathbf{U}(t-1) \Big)^\top.
 \end{align*}
 The local network flow vector for node $i$ up to  time $t$ is defined as
\begin{align*}
 \Theta_i(t):= \Big(\mathbf{s}_j(0)^\top,\dots, \mathbf{s}_j(t)^\top;  \mathbf{u}_j(0),\dots,\mathbf{u}_j(t-1): j\in\mathrm{N}_i \mcup \{i\}\Big)^\top.
 \end{align*}
\end{definition}

Note that, here we have assumed that  the $\mathbf{s}_i(t)$  and  $\mathbf{u}_i(t)$ are known to a node $i$ even if it does not hold a self arc $(i,i)\in\mathrm{E}$ (therefore $i\notin \mathrm{N}_i$).  This is indeed quite natural and general which simplifies the presentation considerably.

\subsubsection{Global-Knowledge/Global-Decision Feedback}

Recall that $A_{\mathrm{G}}$ is the adjacency matrix  of the graph $\mathrm{G}$.
Network controllers that have omniscient narration and omnipotent actuators at all nodes are certainly of primary interest.

\begin{definition} \label{GICDdefinition} A  network control rule  in the form of
\begin{align}\label{GICD}
\mathbf{U}(t) = \mathpzc{h}_t \Big( \Theta(t); A_{\mathrm{G}} \Big),\quad t=0, 1, \dots
\end{align}
where  $\mathpzc{h}_t$ is an arbitrary  function mapping from $\mathbb{R}^{n(3t+1)}$ to $\mathbb{R}^n$ with $A_{\mathrm{G}}$ being a common knowledge, is termed a Global-Knowledge/Global-Decision   Feedback  Law for the network system (\ref{eqn:evol}).
\end{definition}

To implement a global-knowledge/global-decision  network control, one requires  a  network  operator who knows the structure of the network (topology and arc weights),   collects  states and signals across the entire network, and then enforces control  decisions on each individual node.
\subsubsection{Network-Flow/Local-Decision Feedback}
  Knowing the network flow, nodes can still carry out individual control decisions even without knowledge of the entire network structure $\mathrm{G}$. This will incur restrictions on feasible control rules, leading to the following definition.
\begin{definition}\label{GIDDdefinition} A network  control rule  in the form of
\begin{align}\label{GIDD}
\mathbf{u}_i(t) = \mathpzc{h}_t^i \Big( \Theta(t);[A_{\mathrm{G}}]_{ij}, j\in \mathrm{N}_i \Big)
\end{align}
where  independent with other nodes, $\mathpzc{h}^i_t$ is an arbitrary  function mapping from $\mathbb{R}^{|\mathrm{N}_i\mcup\{i\}|(3t+1)}$ to $\mathbb{R}$ for any $t=0,1,\dots$, is termed a Network-Flow/Local-Decision Feedback Law for the network system (\ref{eqn:evol}).
\end{definition}

The $\mathpzc{h}^i_t$ being independent means that a node $m$ can determine its control rule $\mathpzc{h}^m_t$ without knowing or influencing the exact control decision values at any other node and for any given time.  The following example helps clarify the ambiguity in the notion of independent decisions.

\medskip

\noindent {\bf Example 1}.  Consider  two nodes $1$ and $2$. The following control rule with $\mathpzc{q}_t$ being a function with proper dimension for its argument
\begin{align}\label{example1}
\begin{split}
\mathbf{u}_1(t)& = \mathpzc{q}_t( \Theta(t))\\
\mathbf{u}_2(t)& = 1- \mathpzc{q}_t( \Theta(t))
\end{split}
\end{align}
implicitly holds the identity
$$
\mathbf{u}_1(t)+\mathbf{u}_2(t)=1
$$
and therefore can only be implemented if the two nodes coordinate their respective inputs. In this sense (\ref{example1}) is a  global-knowledge/global-decision  feedback rather than a network-flow/local-decision feedback law. \hfill$\square$

\subsubsection{Local-Flow/Local-Decision Feedback}
The notion of distributed control consists of three basis elements \cite{Magnus-Book-2010}: nodes only have a local knowledge of the network structure; nodes only receive and send information to a few neighbors; control and decision are computed by each node independently. Inspired by these criteria we impose the following definition.

\begin{definition} \label{LIDDdefinition}
Any feedback control rule in the form of
\begin{align}\label{distributedlaw}
\mathbf{u}_i(t) = \mathpzc{h}^{i}_{t} \Big(\Theta_i(t); [A_{\mathrm{G}}]_{ij}, j\in \mathrm{N}_i\Big)
\end{align}
with $\mathpzc{h}^i_t: \mathbb{R}^{|\mathrm{N}_i\mcup\{i\}|(3t+1)} \rightarrow \mathbb{R}$ being an arbitrary   function independent with other nodes, is termed a  Local-Flow/Local-Decision  Feedback Law for the network system (\ref{eqn:evol}).
\end{definition}

The three classes of network feedback laws are certainly not disjoint. In fact the set of global-knowledge\dc{}/global\dc{}-decision feedback contains the set of network-flow/local-decision feedback, which in turn contains the set of local-flow/local-decision feedback.

\subsection{Network Stabilizability}
We are interested in the existence of feedback control laws that stabilize the network dynamics  (\ref{eqn:evol}) for the closed loop, as indicated in the following definition.
\begin{definition}
A  feedback law stabilizes the network dynamics  (\ref{eqn:evol}) if there holds
\begin{align}\label{eq:definition_stability}
\sup_{t\geq 0}\Big( \big|\mathbf{x}_i(t) \big|+ \big|\mathbf{z}_i(t) \big|+\big|\mathbf{u}_i(t) \big|\Big)< \infty, \   i\in\mathrm{V}
\end{align}for the closed loop system.
\end{definition}
%It is easy to see that the stabilizing condition
% $
%\sup_{t\geq 0}(|\mathbf{x}_i(t)|+|\mathbf{u}_i(t)|)< \infty
%$ for all $i\in\mathrm{V}$
% is simply equivalent to
% $$
%  \sup_{t\geq 0}|\mathbf{x}_i(t)|<\infty,\ i\in\mathrm{V}.
%  $$

%\subsection{Discussions}\label{sec:discu}
%In this subsection, we provide some additional discussions on suitable function space for the node dynamical mode $f$ and the possible mechanism of nodes obtaining the estimates $\mathbf{z}_i(t)$.
\subsection{Function Space}
We need a metric quantifying the uncertainty in the node dynamical mode $f$. Let  $\mathpzc{F}$ denote the  space that contains   all $\mathbb{R}\rightarrow\mathbb{R}$ functions, where the $f\in\mathpzc{F}$ are equipped with a quasi-norm defined by 
\[
\|f\|_{\rm q} := \lim_{\alpha\rightarrow\infty} \sup_{x,y\in \mathbb{R}}\frac{|f(x)-f(y)|}{|x-y|+\alpha}.
\]
We refer to \cite{xie-guo2000} for a thorough explanation  of this quasi-norm and the resulting  function space $\mathpzc{F}$. Define    $$
\mathpzc{F}_{L}:=\{f\in \mathpzc{F}:\|f\|_q\leq L\}
$$ as a subspace in $\mathpzc{F}$ consisting of functions    bounded by  $L>0$ under   quasi-norm $\|\cdot\|_{\rm q}$. Functions in $\mathpzc{F}_{L}$ can certainly be discontinuous, but they are closely related to  Lipschitz continuous functions.   The following lemma holds, whose proof can be found in \cite{xie-guo2000}.
\begin{lemma}\label{lemma-function}
 Let $\|f\|_{\rm q}\leq L$. Then for any $\eta>0$, there exists $c\geq 0$ such that \begin{align}\label{eqn:f-function}
|f(x)-f(y)|\leq (L+\eta)|x-y|+c, \ \forall x, y\in \mathbb{R}.
\end{align}
\end{lemma}

As a result of Lemma \ref{lemma-function}, as long as $\|f\|_{\rm q}$ admits a finite number, the stabilizability  condition (\ref{eq:definition_stability}) is equivalent to  $$
\sup_{t\geq 0}(|\mathbf{x}_i(t)|+|\mathbf{u}_i(t)|)< \infty, \ i\in\mathrm{V},
$$ which is in turn  equivalent to
 $$
 \sup_{t\geq 0}|\mathbf{x}_i(t)|<\infty,\ i\in\mathrm{V}.
 $$
Moreover, from  Lemma \ref{lemma-function},  the set $\Gamma_L(f):=\big\{(\eta,c):  \mbox{Eq. (\ref{eqn:f-function}) holds}\big\}
$ is nonempty for any  $f\in \mathpzc{F}_{L}$. We further define a constant $ {W}_f(r)$ associated with any $f\in \mathpzc{F}_{L}$ and $r>L$
\begin{align}\label{eqn:function-bound}
 {W}_f(r):= \inf \big\{c: L+\eta <r, (\eta, c)\in \Gamma_L(f) \big\}.
\end{align}

\section{Network Stabilizability Theorems}\label{sec:results}
In this section, we present a series of possibility and/or impossibility results for the stabilizability of the network dynamics  (\ref{eqn:evol}) for the three categories of feedback laws.

%Then $f^*\Big(\big({3}/{2}+\sqrt{2}\big)/\|A_{\mathrm{G}}\|_{\infty}\Big)$

%
%As it was shown in \cite{xie-guo2000}, for any $f\in \mathpzc{F}_{L}$ with $L <({3}/{2}+\sqrt{2})/\|A_{\mathrm{G}}\|_{\infty}$, the set
%\[
%\mathscr{P}_{\mathrm{G}}(f):=\Big\{(\eta,c): 0<\eta<\big({3}/{2}+\sqrt{2}\big)/\|A_{\mathrm{G}}\|_{\infty} -L, |f(x)-f(y)|\leq (L+\eta)|x-y|+c, \forall x, y\in \mathbb{R}\Big\}
%\]
%is nonempty. In fact, for any $0<\eta<\big({3}/{2}+\sqrt{2}\big)/\|A_{\mathrm{G}}\|_{\infty} -L$, there exists a real number $c$ such that $(\eta,c)\in \mathscr{P}_{\mathrm{G}}(f)$.

\subsection{Global-Knowledge/Global-Decision Feedback}
With global-knowledge/global-decision  feedback, it turns out that,   the infinity norm $\|A_{\mathrm{G}}\|_{\infty}$ of the the adjacency matrix  $A_{\mathrm{G}}$, i.e.,
\begin{align*}
\|A_{\mathrm{G}}\|_{\infty}=\max\limits_{i\in \mathrm{V}}\sum\limits_{j\in\mathrm{N}_i} \big|[A_{\mathrm{G}}]_{ij}\big|
\end{align*}
plays a critical role.

Recall that   $ {W}_f(\cdot)$ is the function defined in (\ref{eqn:function-bound}).  The following theorem characterizes a generic fundamental limit for the capacity of global-knowledge/global-decision  feedback laws.

\begin{theorem}[Generic Fundamental Limit]\label{thm:decent}
 Consider $\mathpzc{F}_L$ in the function space $\mathpzc{F}$. Then there exists a generic  \dc{}Global-\dc{}Know\dc{}ledge\dc{}/\dc{}Global\dc{}-Decision feedback law  that stabilizes the network dynamics (\ref{eqn:evol}) if and only if
$$
L<({3}/{2}+\sqrt{2})/\|A_{\mathrm{G}}\|_{\infty}.
$$
To be precise,
the following statements hold.
\begin{itemize}
\item[(i)] If $L<({3}/{2}+\sqrt{2})/\|A_{\mathrm{G}}\|_{\infty}$, then there exists a  global-knowledge/global-decision   feedback control law that stabilizes the system (\ref{eqn:evol}) for all $f \in  \mathpzc{F}_L$ and for all interaction graphs $\mathrm{G}$.
In fact, with $L<({3}/{2}+\sqrt{2})/\|A_{\mathrm{G}}\|_{\infty}$ we can find a global-knowledge/global-decision   feedback control law that ensures
\[
\limsup\limits_{t\rightarrow\infty}|\mathbf{x}_i(t)|\leq \Big(  {W}_f\big({3}/{2}+\sqrt{2})/\|A_{\mathrm{G}}\|_{\infty} \big)+2e_\ast\Big)\|A_{\mathrm{G}}\|_{\infty}+w_\ast, \ \forall i\in\mathrm{V}.
\]
\item[(ii)]  If $L\geq ({3}/{2}+\sqrt{2})/\|A_{\mathrm{G}}\|_{\infty}$, then for any  global-knowledge/global-decision  feedback law (\ref{GICD}) and any initial value $\mathbf{X}(0)$, there  exist an interaction graph $\mathrm{G}$ and a function  $f\in \mathpzc{F}_L$ under which there holds
    \[
\limsup\limits_{t\rightarrow\infty} \max_{i\in \mathrm{V}}|\mathbf{x}_i(t)|=\infty.
\]

\end{itemize}
\end{theorem}

Note that the error bound of the internal state $\mathbf{x}_i(t)$ in statement (i) can be extended to the external state $\mathbf{z}_i(t)$ by 
\[
\limsup\limits_{t\rightarrow\infty}|\mathbf{z}_i(t)| \leq |f(0)|+(5/2+\sqrt{2})\left( {W}_f\big({3}/{2}+\sqrt{2})/\|A_{\mathrm{G}}\|_{\infty} \big)+2e_\ast\right)+w_\ast(3/2+\sqrt{2})/\|A_{\mathrm{G}}\|_{\infty}.
\]
utilizing the fact that $L<({3}/{2}+\sqrt{2})/\|A_{\mathrm{G}}\|_{\infty}$. Moreover, we should emphasize that the critical value  $
L<({3}/{2}+\sqrt{2})/\|A_{\mathrm{G}}\|_{\infty}
$ established in Theorem \ref{thm:decent} is for general interaction graphs. In fact, as will be shown in its proof, the graph constructed for the necessity proof is a very special one containing exactly one self arc. For a given graph $\mathrm{G}$, e.g., a complete graph or a directed cycle, it is certainly possible that the corresponding network dynamics are stabilizable even with $L\geq({3}/{2}+\sqrt{2})/\|A_{\mathrm{G}}\|_{\infty}$. Finding such feedback capacity values for any given interaction graph seems to be rather challenging, as illustrated in the following example.

\medskip

\begin{figure}
\centering
\includegraphics[width=3.2in]{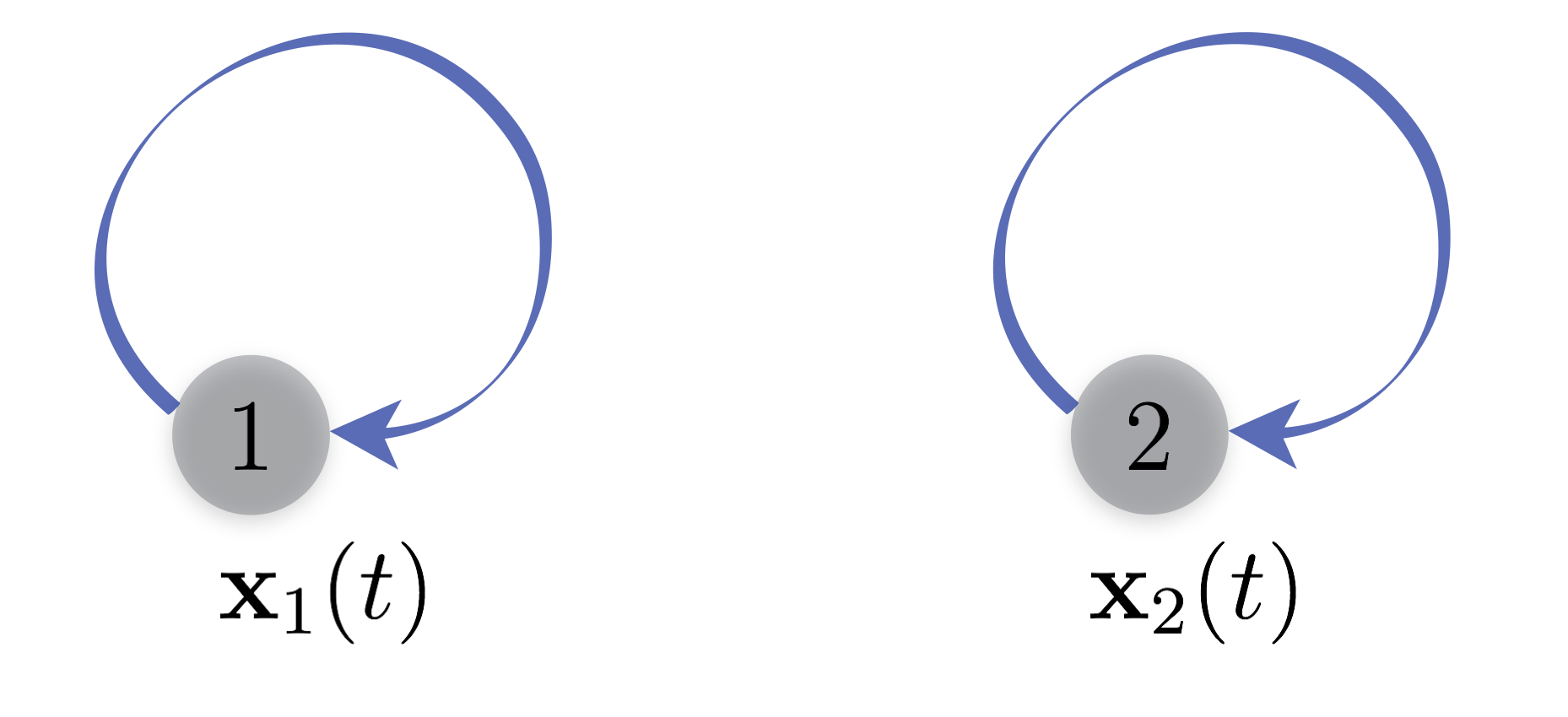}
\caption{A simple two-node network with two self links only.  }
\label{fig:2nodeexample}
\end{figure}

\noindent{\bf Example 2.} Consider two nodes, indexed by $1$ and $2$, respectively,   which both possess a self link with unit weight and have  no link between them (see Fig. \ref{fig:2nodeexample}). From our standing network model their internal  dynamics read as
\begin{align}\label{example2}
\begin{split}
\mathbf{x}_1(t+1)& = f(\mathbf{x}_1(t)) + \mathbf{d}_1(t)+\mathbf{u}_1(t)\\
\mathbf{x}_2(t+1)& = f(\mathbf{x}_2(t)) + \mathbf{d}_2(t)+\mathbf{u}_2(t).
\end{split}
\end{align}
A first sight indicates that (\ref{example2}) appears to be merely two copies of the scaler model considered in \cite{xie-guo2000}. Indeed, directly  from results established in \cite{xie-guo2000}, we know that  if $f\in \mathpzc{F}_L$ with $L<(3/2+\sqrt{2})$, we can stabilize each $\mathbf{x}_i(t)$  with control input $\mathbf{u}_i(t)$ being a feedback from its own dynamics. However, note that with global information, one cannot rule out the case where
 \begin{itemize}
\item[(i)] Node $1$ stabilizes itself;

\item[(ii)] Node $2$   uses the information flow vector\footnote{Node that $\mathbf{z}_1(t)$ can simply be chosen as $\mathbf{x}_1(t+1)-\mathbf{u}_1(t)$.} at the node~$1$:
\begin{align*}
 \Theta_1^\ast(t):= \Big(\mathbf{s}_1(0)^\top,\dots, \mathbf{s}_1(t)^\top;  \mathbf{u}_1(0),\dots,\mathbf{u}_1(t-1)\Big)^\top
 \end{align*}
 to design its controller.
 \end{itemize}
In fact,  $\Theta_1^\ast(t)$ can be rather informative even for node $2$ because it can be utilized putting an effective estimate to the unknown function $f(\cdot)$, which is essential for $\mathbf{u}_2(t)$. Furthermore, one cannot  rule out an even more interesting scenario  where nodes $1$ and $2$ design their controllers {\it cooperatively} since now they share a common   information set. Therefore, it is not clear whether the critical feedback capacity value $3/2+\sqrt{2}$, which applies to the two nodes respectively when they are separate  \cite{xie-guo2000}, will continue to apply  when they form a network with shared  information. An intuitive way of understanding this  is that while the two nodes in system (\ref{example2}) share no {\it dynamical} interaction, a global view of the network information flow will create hidden {\it intellectual}  interaction through their control inputs.    \hfill$\square$

\medskip

Furthermore, we introduce
$$
\|A_{\mathrm{G}}\|_\sharp=\min_{i,j\in\mathrm{V}} \Big\{ \big| [A_{\mathrm{G}}]_{ij}\big|: [A_{\mathrm{G}}]_{ij} \neq 0\Big\}
$$
where  of course $\|A_{\mathrm{G}}\|_\sharp =0$ if $A_{\mathrm{G}}=0$. It is easy to verify that $\|\cdot\|_\sharp$ is not even a proper matrix semi-norm.
The following result however  provides a further impossibility characterization of global-knowledge/global-decision  feedback laws for networks with strong connectivity based on the metric $\|A_{\mathrm{G}}\|_\sharp$.

\begin{theorem}[Impossibility Theorem with Connectivity]\label{thm:impossibility}
Suppose the underlying graph $\mathrm{G}$  is strongly connected. Assume that either  $[A_{\mathrm{G}}]_{ij}\geq 0$ for all $i,j\in\mathrm{V}$ or $[A_{\mathrm{G}}]_{ij}\leq 0$ for all $i,j\in\mathrm{V}$.  If $L \geq 4/\|A_{\mathrm{G}}\|_\sharp$, then for any Global-Knowledge/Global-Decision Feedback Law (\ref{GICD}) and any initial value $\mathbf{X}(0)$, there  exists a  function  $f\in \mathpzc{F}_L$ under which there always holds
    \[
\limsup\limits_{t\rightarrow\infty}|\mathbf{x}_i(t)|=\infty.
\]

\end{theorem}

\subsection{Network-Flow/Local-Decision Feedback}
It is obvious from its definition that any network-flow/local-decision feedback law is by itself  a   global-knowledge/global-decision control as well. In other words, any possibility result for network stabilization achieved by network-flow/local-decision feedback laws can also be viewed as a possibility result for global-knowledge/global-decision controls. Remarkably enough, the contrary also holds true for generic graphs, as indicated in the following result.

\begin{theorem}[Generic Fundamental Limit]\label{thm:GIDD}
 Consider $\mathpzc{F}_L$ in the function space $\mathpzc{F}$. Then there exists a generic \dc{}Network-\dc{}Flow/\dc{}Local-Decision Feedback Law  that stabilizes the network dynamics (\ref{eqn:evol}) if and only if
$
L<({3}/{2}+\sqrt{2})/\|A_{\mathrm{G}}\|_{\infty}.
$
\end{theorem}

In fact, the error bound in Theorem \ref{thm:decent}.(i) continues to hold for network-flow/local-decision feedback laws. Putting Theorem \ref{thm:decent} and Theorem \ref{thm:GIDD} together we learn that, for generic interaction graphs, information flow plays a more critical role for feedback capacity compared to decision structures.

\subsection{Local-Flow/Local-Decision Feedback}
Recall that $a_{ij}=[A_{\mathrm{G}}]_{ij}$ is the weight of arc $(j,i)\in\mathrm{E}$. Let $\langle p^i_t\rangle_{t=1}^\infty$ and  $\langle q^i_t\rangle_{t=1}^\infty$ be non-negative sequences for $i\in\mathrm{V}$ that satisfy the following recursive relations:
\begin{align}\label{inequality}
\begin{split}
p^i_{t+1}  &\leq  \Big(M\sum\limits_{j\in\mathrm{N}_i} |a_{ij}|\max\limits_{1\leq s\leq t}\{p_s^{j},q_s^{j}\} +\omega-\sum\limits_{s=1}^t p^i_{s}\Big) ^+, \\
q^i_{t+1}  &\leq  \Big(M\sum\limits_{j\in\mathrm{N}_i} |a_{ij}|\max\limits_{1\leq s\leq t}\{p_s^{j},q_s^{j}\} +\omega-\sum\limits_{s=1}^t q^i_{s}\Big) ^+.
\end{split}
\end{align}
Induced by recursion (\ref{inequality}), we present  the following metric for the matrix $A_{\mathrm{G}}$
\begin{align}
\|A_{\mathrm{G}}\|_{\dag}:= \sup\Big\{M: \mbox{For any } \omega>0\ {\rm Eq.} (\ref{inequality})\ \mbox{implies}\ \sum_{t=1}^\infty (p^i_t+q^i_t)<\infty \ \mbox{for all } i\in \mathrm{V}\Big\}.
\end{align}
Note that the positivity of  $\|A_{\mathrm{G}}\|_{\dag}$ can be shown for nontrivial graphs $\mathrm{G}$ by establishing $\|A_{\mathrm{G}}\|_{\dag}\geq 1/\|A_{\mathrm{G}}\|_{\infty}$. This observation enabling that $\|A_{\mathrm{G}}\|_{\dag}$ be a meaningful metric for the graph $\mathrm{G}$ has been  put in Lemma \ref{lemma:positivity} as  Appendix.

The following theorem establishes a sufficiency condition for feedback stabilizability of the network dynamics, effectively providing a lower bound of the feedback capacity for  \dc{}local\dc{}-flow\dc{}/\dc{}local-decision feedback laws.
\begin{theorem}[Generic  Possibility Theorem]\label{thm:LIDD}
 Consider $\mathpzc{F}_L$ in the function space $\mathpzc{F}$.  There exists a generic \dc{}Local\dc{}-Flow\dc{}/\dc{}Local-Decision Feedback Law  that stabilizes the network dynamics (\ref{eqn:evol}) if
$$
L/\|A_{\mathrm{G}}\|_{\dag}<1.
$$ More precisely, if $
L<  \|A_{\mathrm{G}}\|_{\dag},
$ then there exists a  \dc{}Local\dc{}-Information\dc{}/\dc{}Local-Decision feedback law that stabilizes the network dynamics (\ref{eqn:evol}) for all $f\in\mathpzc{F}_L$ and all graphs $\mathrm{G}$.
\end{theorem}
\subsection{Max-Consensus Enhanced Feedback Capacity}
It is evident from the above discussions that knowledge of  information flows heavily influences the capacity of feedback laws. Network flow enables universal feedback laws that apply to generic graphs as shown in Theorem \ref{thm:decent} and Theorem \ref{thm:GIDD}, while    local flows can be rather insufficient in stabilizing a network with uncertainty.

However, various distributed algorithms have been developed in the literature serving the aim of achieving collective goals using local node interactions only, which often leads to propagation of certain  global information to local levels. One particular type of such algorithms is the so-called {\it max-consensus}, where a network of nodes holding real values can agree on the network maximal value in finite time steps by distributed interactions \cite{saber2004,jadbabaie2006}. Max-consensus algorithms themselves have been adapted to various settings in complex networks \cite{Giannini2013}, and have been applied to many engineering problems such as sensor network synchronization \cite{Iutzeler2012}. In this subsection, we show simple max-consensus algorithms can fundamentally change the nature of  network feedback capacity.

\medskip

\noindent[{\bf Max-Consensus Enhancement}] At time $t$, each node $i$ holds a vector $\mathbf{s}_i(t)=(\mathbf{x}_i(t), \mathbf{z}_i(t))^\top$. From time $t$ to $(t+1)^-$, nodes run a max-consensus algorithm on the first entry by
$$
\mathbf{s}_i[k+1]=(\mathbf{x}_{\arg \max_{j\in\mathrm{N}_i} \mathbf{x}_j [k]}, \mathbf{z}_{\arg \max_{j\in\mathrm{N}_i} \mathbf{x}_j[k]})^\top
$$
where with slight abuse of notation we neglect the time index $t$ in $\mathbf{m}_i$, $\mathbf{x}_i$, and $\mathbf{z}_i$, and use $[k]$ to represent time steps in the max-consensus algorithm. It is clear \cite{jadbabaie2006} that in a finite number of steps in $k$ (therefore it is safe to assume before time $t+1$), all nodes will  hold
$$
\overline{\mathbf{s}}(t)=\big(\overline{\mathbf{x}}(t), \overline{\mathbf{z}}(t))^\top
$$
with $\overline{\mathbf{x}}(t)=\max_i \mathbf{x}_i(t)$ and   $\overline{\mathbf{z}}(t)= \mathbf{z}_{\arg \max_{j\in\mathrm{V}} \mathbf{x}_j(t)}(t)$.

\medskip

Similarly, $
\underline{\mathbf{s}}(t)=\big(\underline{\mathbf{x}}(t), \underline{\mathbf{z}}(t)\big)^\top
$
with $\underline{\mathbf{x}}(t)=\min_i \mathbf{x}_i(t)$ and   $\underline{\mathbf{z}}(t)= \mathbf{z}_{\arg \min_{j\in\mathrm{V}} \mathbf{x}_j(t)}(t)$ can also be possessed by all nodes $i$ before time $t+1$ with another parallel min-consensus algorithm. We are now ready to introduce the following definition.

\begin{definition}\label{def:Maxenhanced}
The max-consensus enhanced local flow vector for node $i$ up to  time $t$ is defined as
\begin{align*}
\Theta_i^{\rm e}(t):= \Big(\Theta_i(t)^\top, \ \overline{\mathbf{s}}(0)^\top,\dots,\overline{\mathbf{s}}(t)^\top,\ \underline{\mathbf{s}}(0)^\top,\dots,\underline{\mathbf{s}}(t)^\top\Big)^\top.
 \end{align*}
 Moreover,  any feedback control rule in the form of
\begin{align}\label{distributedlaw}
\mathbf{u}_i(t) = \mathpzc{h}^{i}_{t} \Big(\Theta_i^{\rm e}(t); [A_{\mathrm{G}}]_{ij}, j\in \mathrm{N}_i\Big)
\end{align}
with $\mathpzc{h}^i_t$ being an arbitrary function independent with other nodes, is termed a Max\dc{}-Enhanced\dc{}-Local-\dc{}Flow/\dc{}Local-Decision  Feedback Law for the network system (\ref{eqn:evol}).
\end{definition}

It turns out that, max-consensus-enhanced-local-flow/local-decision  feedback laws have the same capacity in stabilizing the generic  network dynamics (\ref{eqn:evol}) as the global-knowledge/global-decision  feedback.

\begin{theorem}[Generic Fundamental Limit]\label{thm:max}
 Consider $\mathpzc{F}_L$ in the function space $\mathpzc{F}$. Then there exists a generic  Max-Consensus-Enhanced-Local-Flow/Local-Decision  Feedback Law  that stabilizes the network dynamics (\ref{eqn:evol}) if and only if
$
L<({3}/{2}+\sqrt{2})/\|A_{\mathrm{G}}\|_{\infty}.
$
\end{theorem}

Although Theorem \ref{thm:max} exhibits the same fundamental limit as Theorem \ref{thm:decent}, the error bound of $\limsup\limits_{t\rightarrow\infty}|\mathbf{x}_i(t)|$ becomes inevitably more conservative. This suggests potential difference at performance levels for the two different types of controllers.

\section{Stabilizing Feedback Laws}\label{sec:controllers}
In this section, we present the control rules that are used in the possibility claims of the above network stabilization   theorems.

\subsection{Local Feedback with Network Flow}
We now present a local feedback controller in the form of Definition \ref{GIDDdefinition} with entire network flow information.  Denote
\begin{align}
\overline{\mathbf{y}}(t) &: = \max\{\mathbf{x}_i(s):s = 0, \dots, t; i = 1,\dots, n\},\label{eqn:overlinex}\\
\underline{\mathbf{y}}(t) &: = \min\{\mathbf{x}_i(s):s = 0, \dots, t; i = 1,\dots, n \}.\label{eqn:underlinex}
\end{align}
as the maximal and minimal states at all nodes and among all time steps up to $t$, respectively.  The controller contains two parts, an estimator and a distributed feedback rule.
\medskip

\noindent [{\bf Estimator}] For each $i\in\mathrm{V}$, $t\geq 1$, there exists  $[v_i]_t\in \mathrm{V}$ and $0\leq [s_i]_t\leq t-1$  that satisfies
\begin{align}\label{eqn:nnxglobal}
\mathbf{x}_{[v_i]_t}([s_i]_t) \in    {\arg\min}_{\mathbf{x}_j(\tau)} \Big\{|\mathbf{x}_{i}(t)-\mathbf{x}_{j}(\tau)|: j\in \mathrm{V}, \tau\in[0,t-1] \Big\}.
\end{align}
Then at time $t$, an estimator for $f(\mathbf{x}_i(t))$ made by nodes that are $i$'s neighbors is given by
\begin{align}\label{eqn:nn-estimator}
\widehat{f}(\mathbf{x}_i(t)):= \mathbf{z}_{[v_i]_t}([s_i]_t+1).
\end{align}

\medskip

\noindent [{\bf Feedback}] Fix any positive $\epsilon$. Let $\mathbf{u}_i(0)=0$ for all $i\in \mathrm{V}$. For all $t\geq 1$ and all $i\in\mathrm{V}$, we define
\begin{equation}\label{eqn:control1}
\mathbf{u}_i(t)=\begin{cases}
 -\sum\limits_{j\in\mathrm{N}_i} a_{ij}\widehat{f}(\mathbf{x}_j(t)), & \text{if } |\mathbf{x}_k(t)-\mathbf{x}_{[v_k]_t}([s_k]_t)|\leq\epsilon \text{\ for all $k\in \mathrm{V}$;} \\
  -\Big(\sum\limits_{j\in\mathrm{N}_i} a_{ij} \widehat{f}(\mathbf{x}_j(t))              \Big) + \frac{1}{2}(\underline{\mathbf{y}}(t)+\overline{\mathbf{y}}(t) ),      & \text{otherwise.}
  \end{cases}
  \end{equation}

It is clear that Eq. (\ref{eqn:nn-estimator})--(\ref{eqn:control1}) lead to a well defined Network-Flow/Local-Decision feedback control law that is consistent with  Definition \ref{GIDDdefinition}. In the following, we will prove that it suffices to use the control law  (\ref{eqn:nn-estimator})--(\ref{eqn:control1}) to establish  the stabilizability statements in Theorem \ref{thm:decent} and Theorem \ref{thm:GIDD}.

\subsection{Global Feedback with Global Information}
\begin{figure*}
\begin{minipage}[t]{0.5\linewidth}
\centering
\includegraphics[width=2.8in]{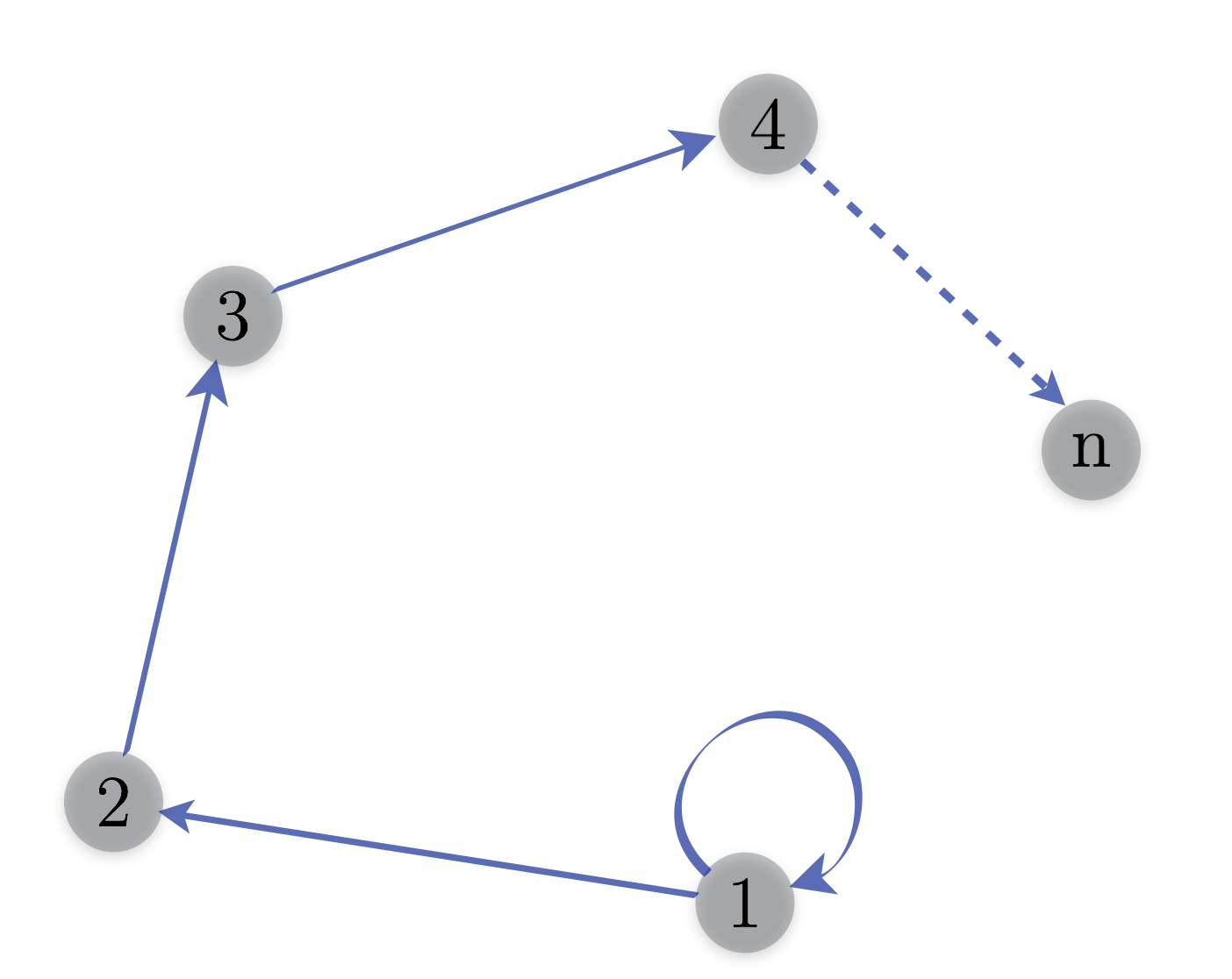}
\end{minipage}%
\begin{minipage}[t]{0.5\linewidth}
\centering
\includegraphics[width=2.8in]{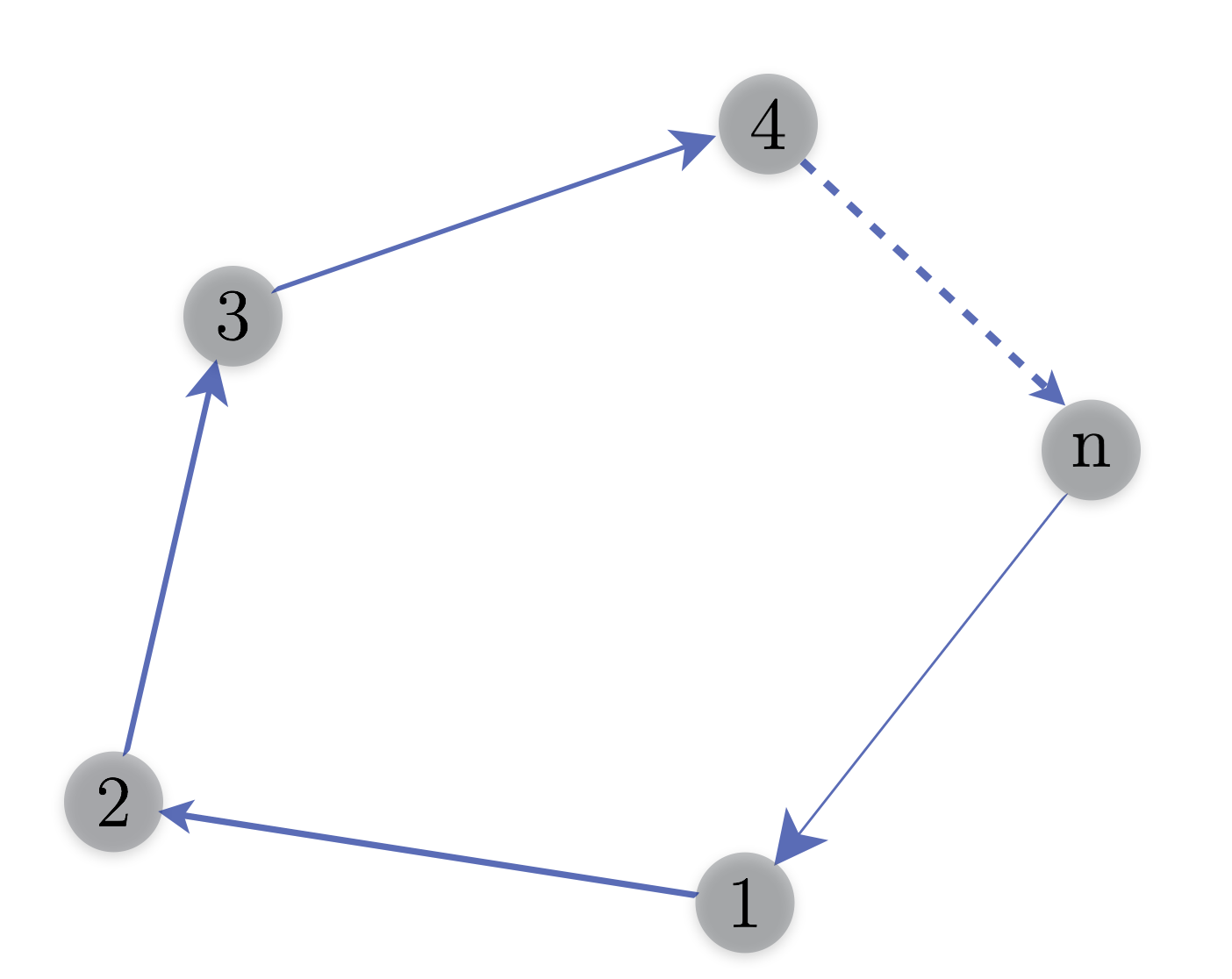}
\end{minipage}
\caption{A directed path graph with one self link at the root node (left), and a directed cycle graph (right). For these two graphs we can construct essential  global-decision controllers that will stabilize the network states.}
\label{fig:path-cycle}
\end{figure*}

The feedback controller given in  (\ref{eqn:nn-estimator})--(\ref{eqn:control1}) already manages to  support the stabilization statement in Theorem \ref{thm:decent} as well since by definition  a local-decision  controller is a special form of global-decision controllers. It is  however of independent interest  seeing  how  stabilizing  network controllers  with essentially centralized structure might work. A clear answer to this question for general graphs seems rather difficult. Nevertheless, we have been able to construct two insightful examples, with the interaction graphs being a directed path  and a directed cycle (see Fig. \ref{fig:path-cycle}), respectively, which partially illustrates some spirit of the problem.

\subsubsection{Path Graph}
Consider the path graph with exactly one self link at the root node\footnote{This self link is added for the sake of providing a  nontrivial example yet as simple as possible.} shown in Fig. \ref{fig:path-cycle} with $a_{11}=1$. Let us consider the following network controller.

\noindent [{\bf Control at root node}]: For each $t\geq 1$, there exists  $0\leq s_t\leq t-1$  that satisfies
\begin{align}\label{eqn:nnxglobal}
\mathbf{x}_{1}(s_t) \in    {\arg\min}_{\mathbf{x}_1(\tau)} \Big\{|\mathbf{x}_{1}(t)-\mathbf{x}_{1}(\tau)|: \tau\in[0,t-1] \Big\}.
\end{align}
At time $t$, an estimator for $f(\mathbf{x}_1(t))$ is given by
\begin{align*}
\widehat{f}(\mathbf{x}_1(t)):= \mathbf{z}_{1}(s_t+1).
\end{align*}
We define
\begin{equation*}
\mathbf{u}_1(t)=
  -\widehat{f}(\mathbf{x}_1(t))+ \frac{1}{2}(\underline{\mathbf{x}}_1(t)+\overline{\mathbf{x}}_1(t) ),
\end{equation*}

\noindent [{\bf Control at other nodes}]: $\mathbf{u}_i(t)=0$ for all $i=2,\dots,n$ and all $t$.

The above network controller will stabilize the node states for any $f\in\mathcal{F}_L$ with $L<{3}/{2}+\sqrt{2}$ citing the result of \cite{xie-guo2000} directly. To implement such a controller, nodes need to know the network structure: node $1$ must know it is a root. All nodes must know $\mathrm{G}$ is a directed path. Nodes $2,\dots,n$ must also know that the controller at node $1$ will stabilize $\mathbf{x}_1(t)$. Therefore, the controller falls into the category of global-knowledge/global-decision network control, but not into  other categories in our definition.
\subsubsection{Cycle Graph}

Consider the directed cycle graph shown in Fig. \ref{fig:path-cycle} and assume all arc weights are equal to one. Define  $\kappa(b)$  for  any (positive, negative, or zero) integer $b\in\mathbb{Z}$ by  $\kappa(b)$ being the unique integer satisfying   $1\leq \kappa(b)\leq n$ and $\kappa(b) =b \mod n$.

%Define $n$ information vectors at time $t$ which will be utilized by each node to design feedback control respectively:
%\begin{align*}
%  \mathscr{I}_1(t) &= \Big(\mathbf{x}_{\kappa(\tau-t)}(\tau): 0\leq \tau \leq t-1\Big)^\top,\\
%  \dots\\
%  \mathscr{I}_i(t) &= \Big(\mathbf{x}_{\kappa(\tau-t+i-1)}(\tau): 0\leq \tau \leq t-1\Big)^\top,\\
%  \dots \\
%  \mathscr{I}_n(t) &= \Big(\mathbf{x}_{\kappa(\tau-t+n-1)}(\tau): 0\leq \tau \leq t-1\Big)^\top.
%\end{align*}
%
%

\medskip

\noindent [{\bf Controller}] For each node $i\in\mathrm{V}$, there exists $0\leq[s_i]_t\leq t-1$ that satisfies
\begin{align}
\mathbf{x}_{\kappa([s_i]_t-t+i-1)}\big([s_i]_t\big)\in{\arg\min}_{\mathbf{x}_{\kappa(\tau-t+i-1)}(\tau)}\Big\{|\mathbf{x}_{\kappa(i-1)}(t)-\mathbf{x}_{\kappa(\tau-t+i-1)}(\tau)|:0\leq\tau\leq t-1\Big\}.
\end{align}
An estimator for $f(\mathbf{x}_{\kappa(i-1)}(t))$ is given by
\begin{align*}
\widehat{f}(\mathbf{x}_{\kappa(i-1)}(t)):= \mathbf{z}_{\kappa([s_i]_t-t+i-1)}\big([s_i]_t+1\big).
\end{align*}
Let $\mathbf{u}_i(0)=0$. For  $t\geq 1$, let
\begin{align}\label{eqn:controlcycle}
  \mathbf{u}_i(t) = - \widehat{f}(\mathbf{x}_{\kappa(i-1)}(t)) + \Big(\max\limits_{0\leq \tau\leq t}\mathbf{x}_{\kappa(\tau-t+i-1)}(\tau)+\min\limits_{0\leq \tau\leq t}\mathbf{x}_{\kappa(\tau-t+i-1)}(\tau)\Big)/2.
\end{align}

\medskip

Clearly  (\ref{eqn:controlcycle}) relies essentially on global decisions because the node number and the cycle structure are necessary knowledge and more importantly, the inherent  symmetry in (\ref{eqn:controlcycle}) requires coordination among the nodes. Suppose  $f\in\mathcal{F}_L$ with $L<{3}/{2}+\sqrt{2}$. Now we show the controller (\ref{eqn:controlcycle}) indeed  stabilizes the  network dynamics.

According to (\ref{eqn:evol}) and the cyclic  network structure, for any $i\in\mathbb{Z}$, there holds
\begin{align}\label{eqn:evolcycle}
\mathbf{x}_{\kappa(i+t+1)}(t+1)= f(\mathbf{x}_{\kappa(i+t)}(t)) + \mathbf{u}_{\kappa(i+t+1)}(t) +\mathbf{d}_{\kappa(i+t+1)}(t).
\end{align}
We further  write $[x_i]_t = \mathbf{x}_{\kappa(i+t)}(t)$, $[d_i]_t = \mathbf{d}_{\kappa(i+t+1)}(t)$, and also $\overline{[x_i]}_t=\max\limits_{0\leq s\leq t}[x_i]_s$, $\underline{[x_i]}_t=\min\limits_{0\leq s\leq t}[x_i]_s$. With these new variables  (\ref{eqn:evolcycle}) becomes
\begin{align}
[x_i]_{t+1} = f([x_i]_t) +\Big(-\widehat{f}([x_i]_t) +\frac{1}{2} (\overline{[x_i]}_t+\underline{[x_i]}_t  )\Big ) +[d_i]_t,
\end{align}
which coincides with the closed loop dynamics for scalar system presented in  \cite{xie-guo2000}. Therefore, quoting the results in \cite{xie-guo2000} we immediately know if $L<{3}/{2}+\sqrt{2}$ then
\begin{align*}
  \limsup\limits_{t\rightarrow\infty}\big|[x_i]_t \big| <\infty, \ i\in\mathrm{V},
\end{align*}
or equivalently, $
  \limsup\limits_{t\rightarrow \infty}|\mathbf{x}_i(t)|<\infty$ and the network dynamics have been  stabilized.

%We could use the controller in \cite{xie-guo2000} to present a simple network controller.
%
%In \cite{xie-guo2000}, the feedback control is provided for the one node case
%\begin{equation}\label{eqn:xieguocontrol}
%  \mathbf{x}_1(t+1) = f (\mathbf{x}_1(t)) +\mathbf{u}_1(t) +\mathbf{w}_1
%\end{equation}
%for $f\in \mathcal{F}_L$ where $L<\frac{3}{2}+\sqrt{2}$. As the definition shows, the feedback control is a function of $(\mathbf{x}_1(0), \dots, \mathbf{x}_1(t);$  $\mathbf{z}_1(0), \dots, \mathbf{z}_1(t-1); $ $\mathbf{u}_1(0), \dots, \mathbf{u}_1(t-1))$. In this case, as discussed in Section \ref{sec:discu}, $\mathbf{z}_1(0), \dots, \mathbf{z}_1(t-1)$ can be obtained by $\mathbf{x}_1(0), \dots, \mathbf{x}_1(t)$. Therefore, the feedback control is determined by on $(\mathbf{x}_1(0), \dots, \mathbf{x}_1(t))$. We denote the feedback control at time $t$ with respect to history information $(\mathbf{x}_1(0), \dots, \mathbf{x}_1(t))$  given in \cite{xie-guo2000} by $\mathpzc{h}^t(\mathbf{x}_1(0), \dots, \mathbf{x}_1(t))$.
%
%We can decompose the control problem on cycle graph as $n$ feedback control problem (\ref{eqn:xieguocontrol}). To write down the control precisely, we introduce some notations. To each $i\in\mathbb{N}$, there exists one and only one $1\leq \kappa(i)\leq n$ such that $\kappa(i) \equiv  i (\rm{mod}$ $ n)$. The feedback control is defined as follows
%\begin{equation}
%  \mathbf{u}_i(t) = \mathpzc{h}^t (\mathbf{x}_{\kappa(i-t)}(0), \mathbf{x}_{\kappa(i-t+1)}(1), \dots \mathbf{x}_{\kappa(i-1)}(t-1),\mathbf{x}_{\kappa(i)}(t)).
%\end{equation}
%

\subsection{Local Feedback with Local Flow}

We now present a local-flow/local-decision feedback law that will enable us to prove Theorem \ref{thm:LIDD}.

\medskip

\noindent [{\bf Estimator}] Fix $i\in\mathrm{V}$. For  $j\in\mathrm{N}_i$ and $t\geq 1$, there exist  $[\mathrm{v}_{ij}]_t\in \mathrm{V}$ and $0\leq [s_{ij}]_t\leq t-1$  that satisfy
\begin{align}\label{eqn:nnxdistributed}
\mathbf{x}_{[\mathrm{v}_{ij}]_t}([s_{ij}]_t) \in    {\arg\min}_{\mathbf{x}_k(s)} \Big\{|\mathbf{x}_{j}(t)-\mathbf{x}_{k}(s)|: k\in \mathrm{N}_i\mcup\{i\}, s\in[0,t-1] \Big\}.
\end{align}
We define an estimator at node $i$ for $f(\mathbf{x}_j(t)),j\in\mathrm{N}_i$ at time $t$ by
\begin{align}\label{eqn:nnestimator-distributed}
 \widehat{f}_i(\mathbf{x}_j(t))=\mathbf{z}_{[\mathrm{v}_{ij}]_t}([s_{ij}]_t+1).
\end{align}

\noindent [{\bf Feedback}] Let $\mathbf{u}_i(0)=0$ for all $i\in \mathrm{V}$. Then for all $t\geq 1$ and all $i\in\mathrm{V}$, we let
\begin{equation}\label{eqn:distributecontrol}
\mathbf{u}_i(t)=
 -\sum\limits_{j\in\mathrm{N}_i} a_{ij} \widehat{f}_i(\mathbf{x}_j(t))+\mathbf{x}_i(0).
\end{equation}

It is also clear that Eq. (\ref{eqn:nnestimator-distributed})--(\ref{eqn:distributecontrol}) form a distributed controller with local information under Definition~\ref{LIDDdefinition}.

\subsection{Local Feedback with Max-Consensus-Enhanced  Local Flow}
Let  $i\in \mathrm{V}$ and  $t\geq 1$.  We denote
\begin{align*}
\mathcal{X}_i(t) & =  \Big\{\overline{\mathbf{x}}(s):0 \leq s\leq t-1\Big\} \mcup \Big\{\underline{\mathbf{x}}(s): 0 \leq s\leq t-1\Big\}\mcup \Big\{\mathbf{x}_j(s): j\in\mathrm{N}_i\mcup\{i\}, 0 \leq s\leq t-1\Big\}
\end{align*}
as the set of states whose estimated data under function $f$ can be accessible to node $i$ at time $t$.
We define a function $\mathcal{K}_i^t(\cdot)$ over $\mathcal{X}_i(t)$ by
\begin{equation}\label{eqn:estimatefxe}
\mathcal{K}_i^t(x) = \left\{
\begin{array}{ll}
\mathbf{z}_j(s+1), & \text{if } x = \mathbf{x}_j(s),\ j\in \mathrm{N}_i\mcup\{i\},\ 0\leq s\leq t-1;\\
\overline{\mathbf{z}}(s+1), & \text{if } x = \overline{\mathbf{x}}(s),\ 0\leq s\leq t-1; \\
\underline{\mathbf{z}}(s+1), & \text{if } x = \underline{\mathbf{x}}(s),\ 0\leq s\leq t-1.
\end{array}
\right.
\end{equation}\label{eqn:nearestnb}

\noindent [{\bf Estimator}] Let node $i$ estimate $f(\mathbf{x}_j(t))$ for $j\in\mathrm{N}_i\mcup \{i\}$ at time $t$ by
\begin{align}\label{eqn:nnestimator-maxenhanced}
 \widehat{f}_i(\mathbf{x}_j(t))=\mathcal{K}_i^t\Big({\arg\min}_{x\in\mathcal{X}_i(t)}\big\{|\mathbf{x}_j(t) -x |\big\} \Big).
\end{align}

\noindent [{\bf Feedback}] Let $\mathbf{u}_i(0)=0$ for all $i\in \mathrm{V}$. Then for all $t\geq 1$ and all $i\in\mathrm{V}$, we let
\begin{equation}\label{eqn:maxenhancedcontrol}
\mathbf{u}_i(t)=
 -\sum\limits_{j\in\mathrm{N}_i} a_{ij} \widehat{f}_i(\mathbf{x}_j(t))+\frac{1}{2}(\overline{\mathbf{y}}(t)+\underline{\mathbf{y}}(t)).
\end{equation}

Eq. (\ref{eqn:nnestimator-maxenhanced})-(\ref{eqn:maxenhancedcontrol}) form a  Max-Consensus-Enhanced-Local-Flow/Local-Decision controller satisfying Definition \ref{def:Maxenhanced}.

\section{Proofs of Statements}\label{sec:proofs}
In this section, we prove  all the claimed    stabilizability theorems.

\subsection{Proof of Theorem \ref{thm:decent} and Theorem \ref{thm:GIDD}}
The sufficiency and necessity of the statements in   Theorem \ref{thm:decent} and Theorem \ref{thm:GIDD} will be proved, respectively, after some helpful technical preparations have been made.

\subsubsection{Preliminary Lemmas}
We first present a few useful  lemmas which turn out incremental for the sufficiency proof, starting from the following technical lemma regarding  two real sequences.
\begin{lemma}\label{lemma:finitesum}
Let $0< M < {3}/{2}+\sqrt{2}$, $t_0\geq 0, \rho\geq 0$ and $\omega$ be any constant. Suppose two nonnegative sequences $\langle p_t\rangle _{t\geq 0}$, $\langle q_t\rangle _{t\geq 0}$ satisfy   for all $t\geq t_0$ that
\begin{align}
p_{t+1} &\leq \Big( M \max\{\max\limits_{1\leq s\leq t} p_{s}, \max\limits_{1\leq s\leq t} q_{s}, \rho\}-\frac{1}{2}\rho-\frac{1}{2}\sum\limits_{s=1}^t (p_s+q_s)    +\omega     \Big) ^+ \label{eqn:h+} \\
q_{t+1} &\leq \Big( M \max\{\max\limits_{1\leq s\leq t} p_{s}, \max\limits_{1\leq s\leq t} q_{s}, \rho\}-\frac{1}{2}\rho-\frac{1}{2}\sum\limits_{s=1}^t (p_s+q_s)    +\omega     \Big) ^+.\label{eqn:g+}
\end{align}
 Then there holds
\begin{equation*}
\sum\limits_{s=1}^{\infty} (p_s+q_s)< \infty.
\end{equation*}
\end{lemma}
{\noindent\it Proof.} We prove the conclusion by contradiction. Assume for the rest of the proof that
\begin{equation}\label{eqn:supposeinfty}
\lim\limits_{t\rightarrow \infty}\sum\limits_{s=1}^{t} (p_s+q_s)=  \infty.
\end{equation}
We divide the argument into a few steps.

{\it\noindent Step 1:} In this step, we prove that for any $t>t_0$, it must hold
\begin{equation}\label{eqn:drop+}
M \max\{\max\limits_{1\leq s\leq t} p_{s}, \max\limits_{1\leq s\leq t} q_{s}, \rho\}-\frac{1}{2}\rho-\frac{1}{2}\sum\limits_{s=1}^t (p_s+q_s)+\omega > 0.
\end{equation}

Let there otherwise exist $t_1>t_0$ such that
\begin{equation*}
 M\max\{\max\limits_{1\leq s\leq t_1} p_{s}, \max\limits_{1\leq s\leq t_1} q_{s}, \rho\}-\frac{1}{2}\rho-\frac{1}{2}\sum\limits_{s=1}^{t_1} (p_s+q_s)+\omega \leq 0.
\end{equation*}
From (\ref{eqn:h+}) and (\ref{eqn:g+}), we immediately know $p_{t_1+1}=0$ and $q_{t_1+1}=0$. This further implies $p_t=0$ and $q_t=0$ for $t>t_1$, which contradicts (\ref{eqn:supposeinfty}). Therefore, (\ref{eqn:drop+}) holds for all $t>t_0$.

{\it\noindent Step 2:} From (\ref{eqn:drop+}),  (\ref{eqn:h+}) and (\ref{eqn:g+}) can be written as
\begin{align}
p_{t+1} & \leq  M \max\{\max\limits_{1\leq s\leq t} p_{s}, \max\limits_{1\leq s\leq t} q_{s}, \rho\}-\frac{1}{2}\rho-\frac{1}{2}\sum\limits_{s=1}^t (p_s+q_s)    +\omega   ,\label{eqn:h}\\
q_{t+1} &\leq M \max\{\max\limits_{1\leq s\leq t} p_{s}, \max\limits_{1\leq s\leq t} q_{s}, \rho\}-\frac{1}{2}\rho-\frac{1}{2}\sum\limits_{s=1}^t (p_s+q_s)    +\omega. \label{eqn:g}
\end{align}
Introduce $\langle r_t\rangle _{t\geq0}$ with $r_0=\rho$ and $
r_t = \max\{p_t, q_t\}$ for $t\geq 1$. There holds for $\langle r_t\rangle $ from (\ref{eqn:h}) and (\ref{eqn:g}) that
\begin{equation}\label{eqn:tildeh}
r_{t+1} \leq M \max\limits_{0\leq s\leq t} r_{s}-\frac{1}{2}\sum\limits_{s=0}^t r_s+\omega.
\end{equation}
In this step, we construct a subsequence $\langle r_{t_m}\rangle _{m\geq 0}$ of $\langle r_{t}\rangle _{t\geq 0}$ with even simpler recursion.

Note that, according to (\ref{eqn:supposeinfty}) and (\ref{eqn:h}), one has
\begin{equation*}
\lim\limits_{t\rightarrow\infty}\max\{\max\limits_{1\leq s\leq t} p_{s}, \max\limits_{1\leq s\leq t} q_{s}, \rho\}=\infty,
\end{equation*}
which yields from the definition of $\langle r_t\rangle $ that
\begin{equation*}
\lim\limits_{t\rightarrow\infty} \max\limits_{0\leq s\leq t} r_{s} = \infty.
\end{equation*}
Thus, there exist subsequence $\langle r_{t_m}\rangle _{m\geq 0}$ of $\langle r_{t}\rangle _{t\geq 0}$ with $r_{t_{m+1}} > r_{t_m}$, such that for $t_m\leq t<t_{m+1}$, there holds $r_{t}\leq r_{t_m}$. As a result, one has $\max\limits_{0\leq s\leq t_{m+1}-1} r_{s}=r_{t_m}$, and further
\begin{equation}\label{eqn:sumtildeh}
r_{t_{m+1}} \leq M r_{t_m} -\frac{1}{2}\sum\limits_{s=0}^{t_{m+1}-1} r_s +\omega\leq M r_{t_m}-\frac{1}{2}\sum\limits_{j=0}^{m} r_{t_{j}}+\omega.
\end{equation}

{\it\noindent Step 3:} This step will conclude the final argument.

Define $R_m = \sum\limits_{j=0}^mr_{t_j}$ for $m \geq 0$, which satisfies $R_{m+1}>R_m$ and $\lim\limits_{m\rightarrow\infty}R_m=\infty$. Replacing $r_{t_k}$ by $R_k-R_{k-1}$ for $k=m, m+1$ in (\ref{eqn:sumtildeh}), we arrive at
\begin{equation}\label{eqn:Hfinal}
R_{m+1} \leq M(R_m-R_{m-1}) + R_m/2+\omega,\ \ \ m\geq 0,
\end{equation}
leading to
\begin{equation}\label{eqn:bk}
\xi_{m+1} \leq M( 1 - {\xi_m^{-1}}) +{1}/{2}+{\omega}/{R_m}, m\geq 1
\end{equation}
where $\xi_m={R_m}/{R_{m-1}}$.
Letting $m\rightarrow\infty$ in (\ref{eqn:bk}) with $\xi:=\liminf\limits_{m\rightarrow \infty} \xi_{m}\geq 1$ results in
\begin{equation*}
\xi\leq M(1-{1}/{\xi})+{1}/{2}.
\end{equation*}
Obviously $\xi\neq 1$ and therefore
\begin{equation*}
M\geq \frac{\xi^2-{\xi}/{2}}{\xi-1}\geq \inf_{\xi>1}\frac{\xi^2-{\xi}/{2}}{\xi-1}={3}/{2}+\sqrt{2}.
\end{equation*}
We have now obtain a contradiction and the desired lemma holds.\hfill$\square$

\medskip

Let  $\mathcal{I}_t:= [\underline{\mathbf{y}}(t), \overline{\mathbf{y}}(t)]$ be the minimal interval containing all node states up to time $t\geq 1$, and then  for $t\geq 1$ introduce $\mathcal{R}_t=(\overline{\mathbf{y}}(t-1), \overline{\mathbf{y}}(t)]$, $ \mathcal{L}_t=[\underline{\mathbf{y}}(t), \underline{\mathbf{y}}(t-1))$.
The length of these intervals is denoted as  $|\mathcal{I}_t|= \overline{\mathbf{y}}(t)-\underline{\mathbf{y}}(t)$,
$| \mathcal{R}_t|=\overline{\mathbf{y}}(t)-\overline{\mathbf{y}}(t-1)$, and $| \mathcal{L}_t|=\underline{\mathbf{y}}(t-1)-\underline{\mathbf{y}}(t)$, respectively.
It is easy to observe that $\mathcal{I}_0$, $ \mathcal{R}_s$, $ \mathcal{L}_s$, $s = 1,\dots, t$, are disjoint sets with
\begin{equation}\label{eqn:decomB}
\mathcal{I}_t = \mathcal{I}_0 \mcup \Big(\mcup_{s=1}^{t}  \mathcal{R}_s \Big)\mcup \Big(\mcup_{s=1}^{t}  \mathcal{L}_s \Big).
\end{equation}
The following lemma holds.

\begin{lemma}\label{lemma-interval} Let $f\in \mathpzc{F}_L$ and consider the closed loop dynamics of the system  (\ref{eqn:evol}) with controller  (\ref{eqn:nn-estimator})-(\ref{eqn:control1}).
Let there exist some $k\in \mathrm{V}$ at time  $t$ such that $|\mathbf{x}_k(t)-\mathbf{x}_{[v_k]_t}([s_k]_t)|>\epsilon$. Then for any $\eta>0$, there exists $E_\ast \geq 0$ such that the  $\big \langle  | \mathcal{R}_t| \big \rangle_{t\geq 1}$ and $\big \langle  | \mathcal{L}_t| \big \rangle_{t\geq 1}$ satisfy recursion
\begin{align}\label{eqn:lemma-intervals}
\begin{split}
| \mathcal{R}_{t+1}|  &\leq  \Big(\|A_{\mathrm{G}}\|_{\infty}(L+\eta)\max\{\max\limits_{1\leq s\leq t}| \mathcal{R}_s|,\max\limits_{1\leq s\leq t}| \mathcal{L}_s|, |\mathcal{I}_0|\}-\frac{1}{2}|\mathcal{I}_0|-\frac{1}{2}\sum\limits_{s=1}^t(| \mathcal{R}_s|+| \mathcal{L}_s|)+E_\ast\Big) ^+ \\
| \mathcal{L}_{t+1}|  &\leq  \Big( \|A_{\mathrm{G}}\|_{\infty}(L+\eta)\max\{\max\limits_{1\leq s\leq t}| \mathcal{R}_s|,\max\limits_{1\leq s\leq t}| \mathcal{L}_s|, |\mathcal{I}_0|\}-\frac{1}{2}|\mathcal{I}_0|-\frac{1}{2}\sum\limits_{s=1}^t(| \mathcal{R}_s|+| \mathcal{L}_s|)+E_\ast\Big) ^+.
\end{split}
\end{align}
\end{lemma}
{\it Proof.} First of all, we establish an unconditional upper bound for $|\mathbf{x}_i(t) - \mathbf{x}_{[v_i]_t}([s_i]_t)|$. From (\ref{eqn:decomB}) there holds
\begin{equation}\label{eqn:sumI}
  |\mathcal{I}_t| = |\mathcal{I}_0| + \sum\limits_{s=1}^{t}( | \mathcal{R}_s| + | \mathcal{L}_s|).
\end{equation}
We investigate two cases, respectively.
\begin{itemize}
\item[(i)] Let $\mathbf{x}_i(t)\notin \mathcal{I}_{t-1}$. Then  obviously there holds
$|\mathbf{x}_i(t)-\mathbf{x}_{[v_i]_t}([s_i]_t)| \leq \max\{| \mathcal{R}_t|,| \mathcal{L}_t|\}$.

\item[(ii)] Let $\mathbf{x}_i(t)\in \mathcal{I}_{t-1}$. Then by (\ref{eqn:decomB}), $\mathbf{x}_i(t)$ must be contained in some $ \mathcal{R}_s$, $ \mathcal{L}_s$ or $\mathcal{I}_0$, giving
\begin{equation*}
|\mathbf{x}_i(t)-\mathbf{x}_{[v_i]_t}([s_i]_t)| \leq \max\{\max\limits_{1\leq s\leq t-1}| \mathcal{R}_s|,\max\limits_{1\leq s\leq t-1}| \mathcal{L}_s|, |\mathcal{I}_0|\}.
\end{equation*}
\end{itemize}
Combining the two cases allows to conclude
\begin{equation}\label{eqn:difference}
|\mathbf{x}_i(t) - \mathbf{x}_{[v_i]_t}([s_i]_t)| \leq \max\{\max\limits_{1\leq s\leq t}| \mathcal{R}_s|,\max\limits_{1\leq s\leq t}| \mathcal{L}_s|, |\mathcal{I}_0|\}
\end{equation}
for all $t\geq 1$.

Next, fix any $\eta>0$ and let $c$ be given in Lemma \ref{lemma-function} since $f\in \mathpzc{F}_L$.  The definition of $| \mathcal{R}_{t+1}|$ implies
\begin{equation*}
\left\{\begin{aligned}
| \mathcal{R}_{t+1}| &= 0, & \text{ if } \max\limits_{1\leq i \leq n} \mathbf{x}_i(t+1) \leq \overline{\mathbf{y}}(t),\\
| \mathcal{R}_{t+1}| &= [\max\limits_{1\leq i \leq n} \mathbf{x}_i(t+1)-\frac{1}{2}(\underline{\mathbf{y}}(t)+\overline{\mathbf{y}}(t) )]-\frac{1}{2}|\mathcal{I}_t|,   & \text{ if } \max\limits_{1\leq i \leq n} \mathbf{x}_i(t+1) > \overline{\mathbf{y}}(t).
\end{aligned}
\right.
\end{equation*}
This allows us to obtain
\begin{align}
| \mathcal{R}_{t+1}| &= \max\{\max\limits_{1\leq i \leq n} \mathbf{x}_i(t+1)-\frac{1}{2}(\underline{\mathbf{y}}(t)+\overline{\mathbf{y}}(t) )-\frac{1}{2}|\mathcal{I}_t|, 0\}\nonumber \\
 &= \Big(\max\limits_{1\leq i \leq n} \mathbf{x}_i(t+1)-\frac{1}{2}(\underline{\mathbf{y}}(t)+\overline{\mathbf{y}}(t) )-\frac{1}{2}|\mathcal{I}_t|\Big)^+ \nonumber\\
&\leq \Big(\max\limits_{1\leq i\leq n}|\mathbf{x}_i(t+1)-\frac{1}{2}(\underline{\mathbf{y}}(t)+\overline{\mathbf{y}}(t) )|-\frac{1}{2}|\mathcal{I}_t|\Big)^+\nonumber\\
 &\leq \Big( \max\limits_{1\leq i\leq n}|\mathbf{x}_i(t+1)-\frac{1}{2}(\underline{\mathbf{y}}(t)+\overline{\mathbf{y}}(t) )|-\frac{1}{2}|\mathcal{I}_0|-\frac{1}{2}\sum\limits_{s=1}^t(| \mathcal{R}_s|+| \mathcal{L}_s|)\Big) ^+,\label{eqn:deltaB+}
\end{align}
where the last inequality is from (\ref{eqn:sumI}).

Finally, plugging the controller  (\ref{eqn:nn-estimator})-(\ref{eqn:control1}) in the network dynamics, we  obtain for the closed loop system that
\begin{align}
&\big|\mathbf{x}_i(t+1)-\frac{1}{2}(\underline{\mathbf{y}}(t)+\overline{\mathbf{y}}(t) )\big|\nonumber\\
 & = \Big|\sum_{j\in\mathrm{N}_i} a_{ij}\left( f(\mathbf{x}_j(t))+\mathbf{e}_j(t)\right) + \mathbf{u}_i(t) +\mathbf{w}_i(t) -\frac{1}{2}(\underline{\mathbf{y}}(t)+\overline{\mathbf{y}}(t) )  \Big|\nonumber \\
& =    \Big|\sum\limits_{j\in\mathrm{N}_i} a_{ij} f(\mathbf{x}_j(t))-\sum\limits_{j\in\mathrm{N}_i} a_{ij} \widehat{f}(\mathbf{x}_j(t))              +\sum\limits_{j\in\mathrm{N}_i} a_{ij} \mathbf{e}_j(t)+\mathbf{w}_i(t)\Big|\nonumber\\
& \leq  \sum\limits_{j\in\mathrm{N}_i} |a_{ij}|\Big( \Big|f(\mathbf{x}_j(t))-f(\mathbf{x}_{[v_j]_t}([s_j]_t)) \Big| + \Big|f(\mathbf{x}_{[v_j]_t}([s_j]_t))-\mathbf{z}_{[v_j]_t}([s_j]_t+1) \Big|\Big) +e_\ast\|A_{\mathrm{G}}\|_{\infty}+w_\ast \nonumber \\
& \leq  \sum\limits_{j\in\mathrm{N}_i} |a_{ij}| \cdot\Big|f(\mathbf{x}_j(t))-f(\mathbf{x}_{[v_j]_t}([s_j]_t)) \Big|+2e_\ast\|A_{\mathrm{G}}\|_{\infty} +w_\ast.
\nonumber \\
&\leq  \|A_{\mathrm{G}}\|_{\infty}(L+\eta)\max\big\{\max\limits_{1\leq s\leq t}| \mathcal{R}_s|,\max\limits_{1\leq s\leq t}| \mathcal{L}_s|, |\mathcal{I}_0|\big\}+E_\ast, \label{eqn:vcenter}
\end{align}
where $E_\ast= (c+2e_\ast)\|A_{\mathrm{G}}\|_{\infty} +w_\ast$,  in the second inequality we have used (\ref{eqn:evol}) and Assumption 2,  and the last inequality is derived by (\ref{eqn:f-function}) and (\ref{eqn:difference}). Combining(\ref{eqn:deltaB+}) and (\ref{eqn:vcenter}) eventually gives us
\begin{align}\label{eqn:finalB+}
| \mathcal{R}_{t+1}|  &\leq  \Big(\|A_{\mathrm{G}}\|_{\infty}(L+\eta)\max\{\max\limits_{1\leq s\leq t}| \mathcal{R}_s|,\max\limits_{1\leq s\leq t}| \mathcal{L}_s|, |\mathcal{I}_0|\}-\frac{1}{2}|\mathcal{I}_0|-\frac{1}{2}\sum\limits_{s=1}^t(| \mathcal{R}_s|+| \mathcal{L}_s|)+E_\ast\Big) ^+.
\end{align}

%by the same analysis, we also have
%\begin{align}\label{eqn:finalB-}
%| \mathcal{L}_{t+1}|  &\leq  \Big( \|A_{\mathrm{G}}\|_{\infty}(L+\eta)\max\{\max\limits_{1\leq s\leq t}| \mathcal{R}_s|,\max\limits_{1\leq s\leq t}| \mathcal{L}_s|, |\mathcal{I}_0|\}-\frac{1}{2}|\mathcal{I}_0|-\frac{1}{2}\sum\limits_{s=1}^t(| \mathcal{R}_s|+| \mathcal{L}_s|)+E_\ast\Big) ^+.
%\end{align}

The inequality about  $|\mathcal{L}_{t+1}|$ can be established using a symmetric analysis. This concludes the proof of the desired lemma. \hfill$\square$

\subsubsection{Proof of Sufficiency}
We are now in a place to prove  the sufficiency part of Theorem \ref{thm:decent} and Theorem \ref{thm:GIDD} by showing the controller presented in Section 4.1 stabilizes the network dynamics.

Fix any $\eta>0$ with  $L+\eta<\big({3}/{2}+\sqrt{2}\big)/\|A_{\mathrm{G}}\|_{\infty}$. Let $c$ be given in Lemma \ref{lemma-function}. The proof is organized into a few steps.

\medskip

{\it\noindent Step 1:} In this step, we prove the following claim.

\medskip

\noindent{Claim.} For any $s_0>0$, there exists $\tau>s_0$ such that
$|\mathbf{x}_{i}(\tau)-\mathbf{x}_{[v_{i}]_{\tau}}([s_{i}]_\tau)|\leq\epsilon$ for all $i\in \mathrm{V}$.

\medskip

Fix $t_0>0$. Suppose for any $t>t_0$, there exists $k\in \mathrm{V}$ (which is dependent on $t$) such that $|\mathbf{x}_{k}(t)-\mathbf{x}_{[v_k]_t}([s_k]_t)|>\epsilon$. Lemma \ref{lemma-interval} implies that (\ref{eqn:lemma-intervals}) holds for all $t>t_0$,  leading to
$\sum\limits_{s=1}^{\infty}(| \mathcal{R}_s|+| \mathcal{L}_s|)<\infty$ if we invoke Lemma \ref{lemma:finitesum}. This further enforces the sequences $\langle \mathbf{x}_i(t)\rangle _{t\geq 0}$ to be bounded for all $i\in \mathrm{V}$, which yields
\begin{align*}
\lim\limits_{t\rightarrow\infty}|\mathbf{x}_i(t)-\mathbf{x}_{[v_i]_t}([s_i]_t)|=0.
\end{align*}
As a result, there however must hold with a large $t$ that
$|\mathbf{x}_{i}(t)-\mathbf{x}_{[v_i]_t}([s_i]_t)|\leq\epsilon$ for all $i\in \mathrm{V}$. Therefore, for any $t_0>0$, we can find $\tau>t_0$ such that $|\mathbf{x}_{i}(\tau)-\mathbf{x}_{[v_i]_\tau}([s_i]_\tau)|\leq\epsilon$ for all $i\in \mathrm{V}$. This proves the desired claim.

\medskip

{\it\noindent Step 2:}  In this step, we prove that  the sequence $\langle \mathbf{x}_i(t)\rangle _{t\geq 0}$ is bounded for all $i\in \mathrm{V}$. Fix an arbitrary $s_0$ and let $\tau(s_0)>s_0$ be the time instant in the above claim. Then
\begin{align}
|\mathbf{x}_i(\tau+1)| & = \Big|\sum_{j\in\mathrm{N}_i} a_{ij} f(\mathbf{x}_j(\tau)) + \mathbf{u}_i(\tau) +\mathbf{d}_i(\tau) \Big|\nonumber \\
& =  \Big |\sum\limits_{j\in\mathrm{N}_i} a_{ij} f(\mathbf{x}_j(\tau))-\sum\limits_{j\in\mathrm{N}_i} a_{ij} \widehat{f}(\mathbf{x}_j(\tau)) +\mathbf{d}_i(\tau) \Big|\nonumber\\
& \leq  \sum\limits_{j\in\mathrm{N}_i} |a_{ij}|\Big(\Big|f(\mathbf{x}_j(\tau))-f(\mathbf{x}_{[v_j]_\tau}([s_j]_\tau))\Big| + \Big|f(\mathbf{x}_{[v_j]_\tau}([s_j]_\tau))-\mathbf{z}_{[v_j]_\tau}([s_j]_\tau)\Big|\Big)+e_\ast\|A_{\mathrm{G}}\|_{\infty} +w_\ast \nonumber\\
&\leq  \|A_{\mathrm{G}}\|_{\infty}(L+\eta)\epsilon +(c+2e_\ast)\|A_{\mathrm{G}}\|_{\infty}+w_\ast \label{eqn:xt+12}
\end{align}
 for all $i\in \mathrm{V}$.

%
% for any $s_0$, we can find $\tau>s_0$, such that for all $i\in \mathrm{V}$,
%$|\mathbf{x}_{i}(\tau)-\mathbf{x}_{[v_i]_\tau}([s_i]_\tau)|\leq\epsilon$. Thus,
%by (\ref{eqn:xt+12}), for all $i\in \mathrm{V}$,
%\begin{equation*}
%|\mathbf{x}_i(\tau+1)| \leq \|A_{\mathrm{G}}\|_{\infty}(L+\eta)\epsilon +(c+D_0)\|A_{\mathrm{G}}\|_{\infty}+w_\ast .
%\end{equation*}
%Therefore, for any $s_0$, the set
%\[
%\{t>s_0: |\mathbf{x}_i(t)| \leq \|A_{\mathrm{G}}\|_{\infty}(L+\eta)\epsilon +(c+D_0)\|A_{\mathrm{G}}\|_{\infty}+w_\ast, i\in \mathrm{V}\}
%\]
%is nonempty.
Therefore, choosing $s_0=0$ we can define
\begin{equation*}
t_0 := \inf \Big\{t>0: |\mathbf{x}_i(t)| \leq\|A_{\mathrm{G}}\|_{\infty}(L+\eta)\epsilon +(c+2e_\ast)\|A_{\mathrm{G}}\|_{\infty}+w_\ast,\ \forall i\in \mathrm{V}\Big\}.
\end{equation*}
Moreover, we can continue to recursively define
\begin{equation*}
t_m := \inf \Big\{t>t_{m-1}: |\mathbf{x}_i(t)| \leq \|A_{\mathrm{G}}\|_{\infty}(L+\eta)\epsilon +(c+2e_\ast)\|A_{\mathrm{G}}\|_{\infty}+w_\ast, \ \forall i\in \mathrm{V}\Big\}.
\end{equation*}
This procedure yields  bounded sequences  $\langle \mathbf{x}_i(t_m)\rangle _{m\geq 0}$ for all   $i\in \mathrm{V}$, and as a result,
\begin{equation*}
\lim\limits_{m\rightarrow\infty}\Big|\mathbf{x}_i(t_m)-\mathbf{x}_{[v_i]_{t_m}}([s_i]_{t_m})\Big|=0, \ i\in \mathrm{V}.
\end{equation*}
In other words, there exists $M\in \mathbb{N}$, such that
\begin{equation*}
|\mathbf{x}_i(t_m)-\mathbf{x}_{[v_i]_{t_m}}([s_i]_{t_m})|\leq\epsilon.
\end{equation*}
 for $m>M$ and for all  $i\in \mathrm{V}$.

However, applying the upper bound in  (\ref{eqn:xt+12}) by replacing $\tau$ with $t_m$, we know  for all $i\in \mathrm{V}$,
\begin{equation}\label{100}
|\mathbf{x}_i(t_m+1)|\leq\|A_{\mathrm{G}}\|_{\infty}(L+\eta)\epsilon +(c+2e_\ast)\|A_{\mathrm{G}}\|_{\infty}+w_\ast.
\end{equation}
By the definition of the $\langle t_m\rangle_{m\geq 0}$ (\ref{100}) ensures
$
t_{m+1} = t_m+1
$ be the only possibility for $m>M$.  This means that we have proved for $t>t_{M+1}$ and all $i\in \mathrm{V}$ that
\begin{equation*}
|\mathbf{x}_i(t)|\leq\|A_{\mathrm{G}}\|_{\infty}(L+\eta)\epsilon +(c+2e_\ast)\|A_{\mathrm{G}}\|_{\infty}+w_\ast.
\end{equation*}

\medskip

{\it\noindent Step 3:} In this step, we  further optimize the upper bound of the node states.  Note that the sequences $\langle \mathbf{x}_i(t)\rangle _{t\geq 0}$,  $i\in \mathrm{V}$ are bounded, elementary properties for bounded real sequences give us
\begin{equation*}
\lim\limits_{t\rightarrow\infty}|\mathbf{x}_i(t)-\mathbf{x}_{[v_i]_{t}}([s_i]_{t})|=0.
\end{equation*}
Thus, for any $\epsilon_\ast<\epsilon$, there exists $s_\ast$ such that for $t>s_\ast$ and all $i\in \mathrm{V}$,
\begin{equation*}
|\mathbf{x}_i(t)-\mathbf{x}_{[v_i]_{t}}([s_i]_{t})|\leq\epsilon_\ast.
\end{equation*}
By the same method as we establish  (\ref{eqn:xt+12}), we further have
\begin{align}\label{101}
|\mathbf{x}_i(t+1)|
&\leq \|A_{\mathrm{G}}\|_{\infty}(L+\eta)\epsilon_\ast +(c+2e_\ast)\|A_{\mathrm{G}}\|_{\infty}+w_\ast, \ \ t>s_\ast, i\in \mathrm{V}
\end{align}
As $\epsilon_\ast$ can be arbitrarily small, (\ref{101}) guarantees
\[
\limsup\limits_{t\rightarrow\infty}|\mathbf{x}_i(t)|\leq (c+2e_\ast)\|A_{\mathrm{G}}\|_{\infty}+w_\ast
\]
for all $ i\in \mathrm{V}$.
This upper bound holds  for any  $c$  given in Lemma \ref{lemma-function} associated with  $\eta>0$ satisfying   $L+\eta<\big({3}/{2}+\sqrt{2}\big)/\|A_{\mathrm{G}}\|_{\infty}$. By the definition of $ {M}(\cdot)$, we can further tighten the bound by
\[
\limsup\limits_{t\rightarrow\infty}|\mathbf{x}_i(t)|\leq \Big(  {W}_f\big({3}/{2}+\sqrt{2})/\|A_{\mathrm{G}}\|_{\infty} \big)+2e_\ast\Big)\|A_{\mathrm{G}}\|_{\infty}+w_\ast, \ \forall i\in\mathrm{V}.
\]
Moreover, 
\begin{align*}
|\mathbf{z}_i(t+1)-f(0)| & \leq |f(\mathbf{x}_i(t))-f(0)| +2e_\ast\\
&\leq (L+\eta)|\mathbf{x}_i(t)| +c +2e_\ast\\
&\leq (5/2+\sqrt{2})\left( {W}_f\big({3}/{2}+\sqrt{2})/\|A_{\mathrm{G}}\|_{\infty} \big)+2e_\ast\right)+w_\ast(3/2+\sqrt{2})/\|A_{\mathrm{G}}\|_{\infty}.
\end{align*}
Therefore,
\[
\limsup\limits_{t\rightarrow\infty}|\mathbf{z}_i(t)| \leq |f(0)|+(5/2+\sqrt{2})\left( {W}_f\big({3}/{2}+\sqrt{2})/\|A_{\mathrm{G}}\|_{\infty} \big)+2e_\ast\right)+w_\ast(3/2+\sqrt{2})/\|A_{\mathrm{G}}\|_{\infty}.
\]
We have now proved the sufficiency statements in Theorem \ref{thm:decent} and Theorem \ref{thm:GIDD}.

\begin{figure}
\centering
\includegraphics[width=3.6in]{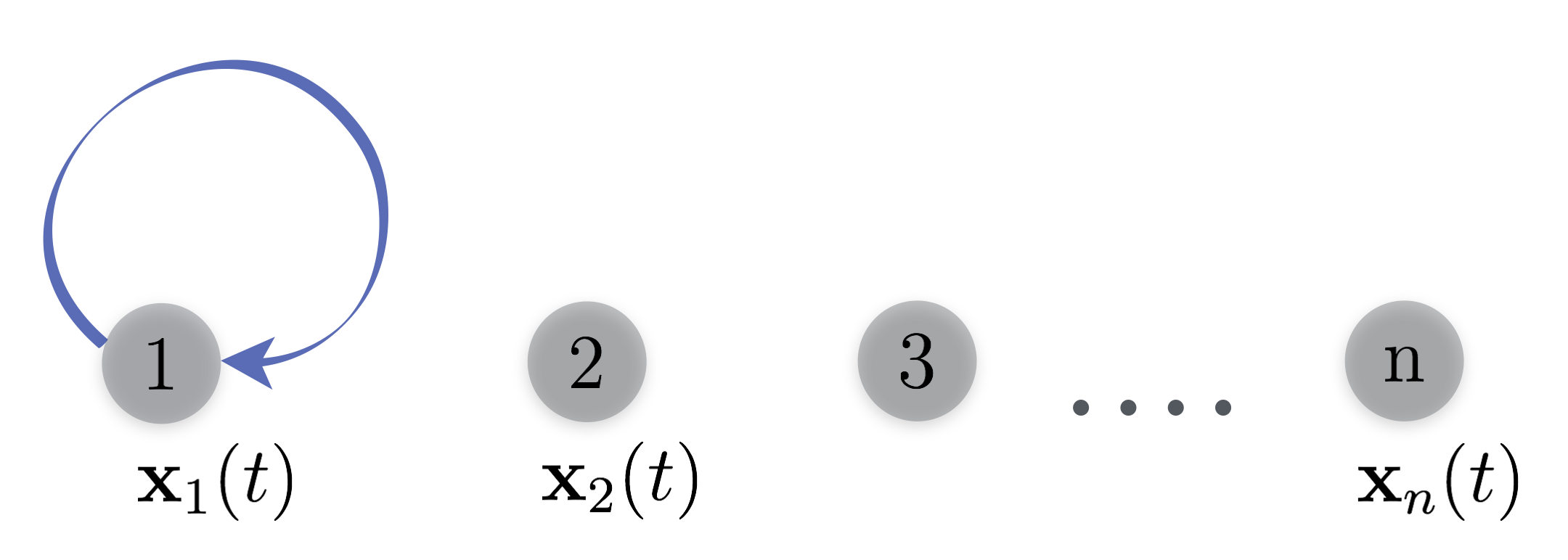}
\caption{An $n$-node network with only one self link.  }
\label{fig:necessity}
\end{figure}

\subsubsection{Proof of Necessity}
Let $L\geq ({3}/{2}+\sqrt{2})/\|A_{\mathrm{G}}\|_{\infty}$ and fix an arbitrary  global-knowledge/global-decision  feedback law (\ref{GICD}).  Fix an initial value $\mathbf{X}(0)$. We continue to construct  an interaction graph $\mathrm{G}$ and a function  $f_\ast \in \mathpzc{F}_L$ under which the network dynamics diverge in the sense that
    \[
\limsup\limits_{t\rightarrow\infty} \max_{i\in\mathrm{V}}|\mathbf{x}_i(t)|=\infty.
\]

The graph we constructed is a simple $n$-node network with only one self link at node $1$ with weight $a_{11}$ (see Fig. \ref{fig:necessity}). Therefore  $\|A_{\mathrm{G}}\|_{\infty}=|a_{11}|$ and the network internal state  dynamics read as
\begin{align}
\begin{split}
\mathbf{x}_1(t+1)&=a_{11}f(\mathbf{x}_1(t))+a_{11}\mathbf{e}_1(t)+\mathbf{w}_1(t)+\mathbf{u}_1(t)\\
\mathbf{x}_2(t+1)&=\mathbf{w}_2(t)+\mathbf{u}_2(t)\\
\vdots \\
\mathbf{x}_n(t+1)&=\mathbf{w}_n(t)+\mathbf{u}_n(t).
\end{split}
\end{align}
Note that, the trajectories of $\mathbf{s}_i(t)$, $t=2,\dots,n$ are by themselves stable with zero inputs, and they contain no information about $f(\cdot)$ regardless of the choice of the $\mathbf{u}_i(t)$ for $t=2,\dots,n$. Therefore, stabilizability of  the above network dynamics is equivalent with stabilizability of the dynamics of node $1$:
\begin{align}\label{99}
\mathbf{x}_1(t+1)&=a_{11}f(\mathbf{x}_1(t))+\mathbf{d}_1(t)+\mathbf{u}_1(t).
\end{align}
This system is essentially the same as the scalar system investigated in \cite{xie-guo2000} except for the known constant $a_{11}$. Invoking the necessity proof of Theorem  2.1 in \cite{xie-guo2000},  we easily know that for any feedback law $\mathbf{u}_1(t)$, we can find a  function $f_\ast \in \mathpzc{F}_L$ with $$
|a_{11}|L=\|A_{\mathrm{G}}\|_{\infty}L= 3/2+\sqrt{2}
$$
under which     the closed loop dynamics of (\ref{99}) lead to
\[
\limsup\limits_{t\rightarrow\infty} |\mathbf{x}_1(t)|=\infty.
\]

We have now concluded the necessity proof of Theorem \ref{thm:decent}, and therefore the necessity proof of Theorem~\ref{thm:GIDD} as well.

\subsection{Proof of Theorem \ref{thm:impossibility}}

\subsubsection{Preliminaries}
Before presenting the main body of the proof we introduce a set of useful concepts and notations.
\begin{definition}
A continuous function $\mathrm{h}:\mathbb{R}\rightarrow\mathbb{R}$ is said to be piecewise linear with slope $\pm B$, if there exists a increasing sequence $\langle y_n\rangle _{n=-\infty}^{\infty}$ with $\lim_{n\rightarrow\infty}y_n=\overline{y} \leq \infty$
and
$\lim_{n\rightarrow-\infty}y_n=\underline{y} \geq -\infty$,
such that $\mathrm{h}$ is linear on $(-\infty, \underline{y}]$, $[y_n, y_{n+1}]$ and $[\overline{y},+\infty)$ with slope $B$ or $-B$ for any $n\in \mathbb{Z}$.
\end{definition}
We denote the set of all piecewise linear functions with slope $\pm B$ as $\mathpzc{H}^*_B$.
It is easy to verify that for any $\mathrm{h}\in \mathpzc{H}^*_B$ and any $x,y\in \mathbb{R}$,
\begin{equation*}
  |\mathrm{h}(x) -\mathrm{h}(y)|\leq B|x-y|.
\end{equation*}
Therefore, $\mathpzc{H}^*_B \subset \mathpzc{F}_B$
for all $B>0$.

We continue to use the notion of intervals $\mathcal{I}_t$, $\mathcal{R}_t$, and $\mathcal{L}_t$ as defined  in the previous proof.  Moreover, for any $t\geq 0$, there exist $[\overline{i}]_t\in \mathrm{V}$ and $0\leq [\overline{s}]_t\leq t$ that satisfy $\mathbf{x}_{[\overline{i}]_t}([\overline{s}]_t) = \overline{\mathbf{y}}(t)$, and  $[\underline{i}]_t\in \mathrm{V}$ and $0\leq [\underline{s}]_t\leq t$ that satisfy $\mathbf{x}_{[\underline{i}]_t}([\underline{s}]_t) = \underline{\mathbf{y}}(t)$. Let $\theta_0=[\overline{i}]_0$ and   for any $t\geq 1$, $  \theta_t = [\overline{i}]_t $ if $| \mathcal{R}_t| \geq | \mathcal{L}_t|$, $  \theta_t = [\underline{i}]_t   $ otherwise.
Since the graph $\mathrm{G}$ is strongly connected, we associate with   $\theta_t\in\mathrm{V}$ an arbitrary  node $d_t\in\mathrm{V}$ that satisfies $(\theta_t,d_t)\in \mathrm{E}$, i.e., $\theta_t$ is a neighbor of $d_t$. We also introduce
\begin{equation}\label{eqn:maxdeltaI+I-}
  \chi(t) = \max\{| \mathcal{R}_t|, | \mathcal{L}_t|\},
\end{equation}
which satisfies trivially from the network dynamics and the definition of $\mathcal{I}(t)$ that
\begin{align} \label{eqn:Rtdistance}
  \chi(t+1) = \max\limits_{1\leq i \leq n} {\rm dist}(\mathbf{x}_i(t+1), \mathcal{I}_t)= {\rm dist}(\mathbf{x}_{\theta_{t+1}}([\overline{s}]_{t+1}), \mathcal{I}_t),
\end{align}
for all $t\geq 0$.

\subsubsection{Construction of the Function}

Let the underlying graph $\mathrm{G}$  be strongly connected. Assume  either $[A_{\mathrm{G}}]_{ij}\geq 0$ for all $i,j\in\mathrm{V}$ or  $[A_{\mathrm{G}}]_{ij}\leq 0$ for all $i,j\in\mathrm{V}$.   Fix the initial value $\mathbf{X}(0)$,  the global-knowledge/global-decision  feedback law $\mathbf{U}(t)$, and the noise function $\mathbf{w}_i(t)$, $\mathbf{e}_i(t)$, $i\in \mathrm{V}$. Thus, the noise functions $\mathbf{d}_i(t)$, $i\in \mathrm{V}$ are given.  We proceed to construct  a  function  $f\in \mathpzc{F}_L$  with $L =4/\|A_{\mathrm{G}}\|_\sharp$ under which the closed-loop network dynamics will  asymptotically diverge. To this end, we first recursively define a serial of function sets.

Denote $\mathpzc{H}^*:=\mathpzc{H}^*_B$ with $B=4/\|A_{\mathrm{G}}\|_\sharp$ as the set of piecewise linear functions with slope $4/\|A_{\mathrm{G}}\|_\sharp$. As mentioned above we have $\mathpzc{H}^*\subseteq \mathpzc{F}_L$  with $L =4/\|A_{\mathrm{G}}\|_\sharp$.

\medskip

{\it\noindent Step 1:} In this step, we define a set of functions on $\mathbb{R}$ that they have common values on interval $\mathcal{I}_0$. Denote $\alpha_0 = \underline{\mathbf{y}}(0)$ and $\alpha_1 = \overline{\mathbf{y}}(0)$, and introduce
\begin{align*}
  \mathpzc{H}^{0}_{\rm p} = \big\{\mathrm{h}\in\mathpzc{H}^*: \mathrm{h}(\alpha_0)=1, \mathrm{h}(\alpha_1) = 1 + {4}|\mathcal{I}_0|/{\|A_{\mathrm{G}}\|_\sharp}\big\},
\end{align*}
and
\begin{align*}
  \mathpzc{H}^{0}_{\rm n} = \big\{\mathrm{h}\in\mathpzc{H}^*: \mathrm{h}(\alpha_0)=-1, \mathrm{h}(\alpha_1) = -1 - {4}|\mathcal{I}_0|/{\|A_{\mathrm{G}}\|_\sharp} \big\}.
\end{align*}
Both sets are nonempty. We define
\begin{equation}\label{eqn:Gpl0}
  \mathpzc{H}^{0}=\left\{\begin{aligned}
  \mathpzc{H}^{0}_{\rm p}, &\quad \text{if } {\rm dist}\Big(\sum\limits_{j\in \mathrm{N}_{d_0}} a_{d_0j} \mathrm{g}_0(\mathbf{x}_j(0)) + \mathbf{u}_{d_0}(0) + \mathbf{d}_{d_0}(0), (\underline{\mathbf{y}}(0)+\overline{\mathbf{y}}(0))/2\Big)\geq\|A_{\mathrm{G}}\|_\sharp +4|\mathcal{I}_0|, \\
  \mathpzc{H}^{0}_{\rm n},      &\quad \text{otherwise,}
  \end{aligned}\right.
\end{equation}
where $\mathrm{g}_0$ is any function in $\mathpzc{H}^{0}_{\rm p}$.
\medskip

{\it\noindent Step 2:} In this step, we define a subset of $\mathpzc{H}^{0}$ that the functions in it hold common values on interval $\mathcal{I}_1$. By the definition of $\mathpzc{H}^{0}$, for any $\mathrm{h}\in \mathpzc{H}^{0}$, $\mathbf{X}(1)$ holds the same value.
As will be shown later in the divergence proof, there holds  $\mathbf{x}_{d_0}(1)\notin \mathcal{I}_0$. Thus, $\mathbf{x}_{\theta_1}([\overline{s}]_1)\notin \mathcal{I}_0$ and $[\overline{s}]_1=1$.
Let $\alpha_2 = \overline{\mathbf{y}}(1)$ and $\alpha_{-1}=\underline{\mathbf{y}}(1)$. We define
\begin{align*}
  \mathpzc{H}_{\rm p}^{1} = \big\{\mathrm{h}\in\mathpzc{H}^{0}: \mathrm{h}(\alpha_2) = \mathrm{h}(\alpha_1) +  {4}| \mathcal{R}_1|/{\|A_{\mathrm{G}}\|_\sharp}   \text{ and } \mathrm{h}(\alpha_{-1}) = \mathrm{h}(\alpha_0) + {4}| \mathcal{L}_1|/{\|A_{\mathrm{G}}\|_\sharp} \big \}
\end{align*}
and
\begin{align*}
  \mathpzc{H}_{\rm n}^{1} = \big \{\mathrm{h}\in\mathpzc{H}^{0}: \mathrm{h}(\alpha_2) = \mathrm{h}(\alpha_1) - {4}| \mathcal{R}_1|/{\|A_{\mathrm{G}}\|_\sharp}  \text{ and } \mathrm{h}(\alpha_{-1}) = \mathrm{h}(\alpha_0) - {4}| \mathcal{L}_1|/{\|A_{\mathrm{G}}\|_\sharp}\big \}.
\end{align*}
Again the two sets are both nonempty. Introduce
\begin{equation}
  \mathpzc{H}^{1}=\left\{\begin{aligned}
  \mathpzc{H}_{\rm p}^{1}, &\quad \text{if } {\rm dist}\Big(\sum_{j\in \mathrm{N}_{d_1}} a_{d_1j} \mathrm{g}_1(\mathbf{x}_j(1)) + \mathbf{u}_{d_1}(1) + \mathbf{d}_{d_1}(1), (\underline{\mathbf{y}}(1)+\overline{\mathbf{y}}(1))/2\Big)\geq 4\chi(1), \\
  \mathpzc{H}_{\rm n}^{1},      &\quad \text{otherwise,}
  \end{aligned}\right.
\end{equation}
where $\mathrm{g}_1$ is any function in $\mathpzc{H}_{\rm p}^{1}$.
\medskip

{\it\noindent Step 3:} In this step, we recursively define the set of functions on $\mathbb{R}$ which take common values on interval $\mathcal{I}_t$. For any $t\geq 2$ and any $\mathrm{h}\in \mathpzc{H}^{t-1}$, $\mathbf{X}(t)$ holds the same value.
There holds that $\mathbf{x}_{d_{t-1}}(t)\notin \mathcal{I}_{t-1}$, whose proof is deferred to the part of divergence analysis. Thus, $\mathbf{x}_{\theta_{t}}([\overline{s}]_t)\notin \mathcal{I}_{t-1}$ and $[\overline{s}]_t=t$. Let $\alpha_{t+1}=\overline{\mathbf{y}}(t)$ and $\alpha_{-t}=\underline{\mathbf{y}}(t)$. Define
\begin{equation*}
  \mathpzc{H}_{\rm p}^{t} = \big\{\mathrm{h}\in\mathpzc{H}^{t-1}: \mathrm{h}(\alpha_{t+1}) = \mathrm{h}(\alpha_{t}) + {4}| \mathcal{R}_t|/\|A_{\mathrm{G}}\|_\sharp \text{ and } \mathrm{h}(\alpha_{-t}) = \mathrm{h}(\alpha_{-t+1}) + {4}| \mathcal{L}_t|/\|A_{\mathrm{G}}\|_\sharp \big \}
\end{equation*}
and
\begin{equation*}
  \mathpzc{H}_{\rm n}^{t} = \big \{\mathrm{h}\in\mathpzc{H}^{t-1}: \mathrm{h}(\alpha_{t+1}) = \mathrm{h}(\alpha_{t}) - {4}| \mathcal{R}_t|/\|A_{\mathrm{G}}\|_\sharp \text{ and } \mathrm{h}(\alpha_{-t}) = \mathrm{h}(\alpha_{-t+1}) - 4| \mathcal{L}_t|/ \|A_{\mathrm{G}}\|_\sharp  \big\}.
\end{equation*}
It is easy to verify they are nonempty sets and further let
\begin{equation}
  \mathpzc{H}^{t}= \begin{cases}
  \mathpzc{H}_{\rm p}^{t}, & \text{if } {\rm dist}\Big(\sum_{j\in \mathrm{N}_{d_t}} a_{d_tj} \mathrm{g}_t(\mathbf{x}_j(t)) + \mathbf{u}_{d_t}(t) + \mathbf{d}_{d_t}(t), (\underline{\mathbf{y}}(t)+\overline{\mathbf{y}}(t))/2\Big)\geq 4\chi(t), \\
  \mathpzc{H}_{\rm n}^{t},      & \text{otherwise,}
  \end{cases}
\end{equation}
where $\mathrm{g}_t$ is any function in $\mathpzc{H}_{\rm p}^{t}$.

\medskip

Finally,  the sequence of functions $\mathpzc{H}^{t}$ specifies   an increasing sequence of real numbers $\big\langle  \alpha_t \big\rangle_{t=-\infty}^{\infty}$. Let
\begin{equation}
    \mathpzc{H}^{\infty} = \big\{\mathrm{h}\in\mathpzc{H}^*: \mathrm{h}(\alpha_{t+1}) = \mathrm{h}_t(\alpha_{t+1}) \text{ and } \mathrm{h}(\alpha_{-t}) = \mathrm{h}_t(\alpha_{-t}) \text{ for any } \mathrm{h}_t\in \mathpzc{H}^{t}, t= 0, 1, \dots  \big\}
\end{equation}
which is certainly not empty. For any given  $f\in \mathpzc{H}^{\infty}$, as will be shown later,  the given  feedback law $\mathbf{U}(t)$ will not be able  stabilize the network dynamics (\ref{eqn:evol}).

\subsubsection{Proof of Divergence}

We now prove that   the  feedback law $\mathbf{U}(t)$ with  the network dynamics (\ref{eqn:evol}) will drive the network dynamics to diverge for any $f_\ast\in \mathpzc{H}^{\infty}$. Our argument is organized in steps as usual.

{\it\noindent Step 1:} We first investigate the network state for $t=1$. Based on the definition of  $\mathpzc{H}^{0}_{\rm p}$ and  $\mathpzc{H}^{0}_{\rm n}$, we know for any $\mathrm{g}_0\in \mathpzc{H}^{0}_{\rm p}$ and $\mathrm{h}_0\in \mathpzc{H}^{0}_{\rm n}$
\begin{align}
   &\quad{\rm dist}\Big(\sum_{j\in \mathrm{N}_{d_0}} a_{d_0j} \mathrm{g}_0(\mathbf{x}_j(0)) + \mathbf{u}_{d_0}(0) + \mathbf{d}_{d_0}(0), \sum_{j\in \mathrm{N}_{d_0}} a_{d_0j} \mathrm{h}_0(\mathbf{x}_j(0)) + \mathbf{u}_{d_0}(0) + \mathbf{d}_{d_0}(0)\Big)\nonumber \\
  &= \Big|\sum_{j\in \mathrm{N}_{d_0}} a_{d_0j} (\mathrm{g}_0(\mathbf{x}_j(0))-\mathrm{h}_0(\mathbf{x}_j(0)) )\Big| \nonumber\\
  &= \sum_{j\in \mathrm{N}_{d_0}} |a_{d_0j}| \cdot \Big |\mathrm{g}_0(\mathbf{x}_j(0))-\mathrm{h}_0(\mathbf{x}_j(0)) \Big| \nonumber\\
  &\geq  |a_{d_0\theta_0}| \cdot \Big|\mathrm{g}_0(\mathbf{x}_{\theta_0}(0))-\mathrm{h}_0(\mathbf{x}_{\theta_0}(0))\Big|\nonumber\\
  &\geq  \|A_{\mathrm{G}}\|_\sharp(2+{8}|\mathcal{I}_0|/{\|A_{\mathrm{G}}\|_\sharp}) \nonumber\\
  &= 2\|A_{\mathrm{G}}\|_\sharp + 8 |\mathcal{I}_0|,
\end{align}
  where the second equality holds due to the fact that all $a_{ij}$ have the same sign for $(i,j)\in\mathrm{E}$. Thus, for any $\mathrm{h}\in\mathpzc{H}^{0}$,
\begin{align}\label{eqn:g0evol}
  {\rm dist}\Big(\sum\limits_{j\in \mathrm{N}_{d_0}} a_{d_0j} \mathrm{h}(\mathbf{x}_j(0)) + \mathbf{u}_{d_0}(0) + \mathbf{d}_{d_0}(0), \frac{1}{2}(\underline{\mathbf{y}}(0)+\overline{\mathbf{y}}(0)\Big) \geq \|A_{\mathrm{G}}\|_\sharp +4|\mathcal{I}_0|.
\end{align}

%Therefore,
%\begin{align}\label{eqn:g0+distance}
%  {\rm dist}\big(\sum_{j\in \mathrm{N}_{d_0}} a_{d_0j} \mathrm{g}_0(\mathbf{x}_j(0)) + \mathbf{u}_{d_0}(0) + \mathbf{w}_{d_0}(0), \frac{1}{2}(\underline{\mathbf{y}}(0)+\overline{\mathbf{y}}(0))\big)\geq \|A_{\mathrm{G}}\|_\sharp +4|\mathcal{I}_0|,
%\end{align}
%or
%\begin{align}\label{eqn:g0-distance}
%{\rm dist}\big(\sum_{j\in \mathrm{N}_{d_0}} a_{d_0j} \mathrm{h}_0(\mathbf{x}_j(0)) + \mathbf{u}_{d_0}(0) + \mathbf{w}_{d_0}(0),\frac{1}{2}(\underline{\mathbf{y}}(0)+\overline{\mathbf{y}}(0))\big)\geq \|A_{\mathrm{G}}\|_\sharp +4|\mathcal{I}_0|.
%\end{align}
%Here, we define
%\begin{equation}\label{eqn:Gpl0}
%  \mathpzc{H}^{0}=\left\{\begin{aligned}
%  \mathpzc{H}^{0}_{\rm p}, &\quad \text{if } {\rm dist}\big(\sum\limits_{j\in \mathrm{N}_{d_0}} a_{d_0j} \mathrm{g}_0(\mathbf{x}_j(0)) + \mathbf{u}_{d_0}(0) + \mathbf{w}_{d_0}(0), \frac{1}{2}(\underline{\mathbf{y}}(0)+\overline{\mathbf{y}}(0))\big)\geq\|A_{\mathrm{G}}\|_\sharp +4|\mathcal{I}_0|, \\
%  \mathpzc{H}^{0}_{\rm n},      &\quad \text{otherwise.}
%  \end{aligned}\right.
%\end{equation}
Now that
\begin{align*}
  \mathbf{x}_{d_0}(1) = \sum_{j\in \mathrm{N}_{d_0}} a_{d_0j} {f}_\ast(\mathbf{x}_j(0)) + \mathbf{u}_{d_0}(0) + \mathbf{d}_{d_0}(t)
\end{align*}
we know
\begin{align}\label{eqn:R1}
  \chi(1) &= {\rm dist}\big((\mathbf{x}_{\theta_1}([\overline{s}]_1), \mathcal{I}_0\big) \nonumber\\
  &\geq {\rm dist}\big((\mathbf{x}_{d_0}(1), \mathcal{I}_0 \big)\nonumber \\
  &\geq {\rm dist}\big(\mathbf{x}_{d_0}(1),(\underline{\mathbf{y}}(0)+\overline{\mathbf{y}}(0))/2\big)-|\mathcal{I}_0|/2\nonumber\\
  &={\rm dist}\Big(\sum_{j\in \mathrm{N}_{d_0}} a_{d_0j} f_\ast(\mathbf{x}_j(0)) + \mathbf{u}_{d_0}(0) + \mathbf{d}_{d_0}(t), \frac{1}{2}(\underline{\mathbf{y}}(0)+\overline{\mathbf{y}}(0) )\Big)-|\mathcal{I}_0|/2\nonumber\\
  &\geq \|A_{\mathrm{G}}\|_\sharp+ 7|\mathcal{I}_0|/2
\end{align}
according to (\ref{eqn:Rtdistance}) and (\ref{eqn:g0evol}) as ${f}_\ast$ coincides with $\mathrm{h}$ on $\mathcal{I}_0$. This also indicate that $\mathbf{x}_{d_0}(1)\notin \mathcal{I}_0$.

{\it\noindent Step 2:} Next, we investigate the case with $t=2$ and reveal the recursion pattern. For any $\mathrm{g}_1\in \mathpzc{H}_{\rm p}^{1} $ and $\mathrm{h}_1\in \mathpzc{H}_{\rm n}^{1}$, one has
\begin{align}
  &\quad {\rm dist}\Big(\sum_{j\in \mathrm{N}_{d_1}} a_{d_1j} \mathrm{g}_1(\mathbf{x}_j(1)) + \mathbf{u}_{d_1}(1) + \mathbf{d}_{d_1}(1), \sum_{j\in \mathrm{N}_{d_1}} a_{d_1j} \mathrm{h}_1(\mathbf{x}_j(1)) + \mathbf{u}_{d_1}(1) + \mathbf{d}_{d_1}(1)\Big)\nonumber \\
  &= \Big|\sum_{j\in \mathrm{N}_{d_1}} a_{d_1j} (\mathrm{g}_1(\mathbf{x}_j(1))-\mathrm{h}_1(\mathbf{x}_j(1)) ) \Big| \nonumber\\
  &= \sum_{j\in \mathrm{N}_{d_1}} |a_{d_1j}| \cdot \Big |\mathrm{g}_1(\mathbf{x}_j(1))-\mathrm{h}_1(\mathbf{x}_j(1)) \Big| \nonumber\\
  &\geq  |a_{d_1\theta_1}|\cdot \big |\mathrm{g}_1(\mathbf{x}_{\theta_1}(1))-\mathrm{h}_1(\mathbf{x}_{\theta_1}(1))\big| \nonumber\\
  &\geq  8 \chi(1). \nonumber
\end{align}
Thus, for any $\mathrm{h}\in\mathpzc{H}^{1}$,
\begin{equation}\label{eqn:g1evol}
  {\rm dist}\Big(\sum\limits_{j\in \mathrm{N}_{d_1}} a_{d_1j} \mathrm{h}(\mathbf{x}_j(1)) + \mathbf{u}_{d_1}(1) + \mathbf{d}_{d_1}(1), \frac{1}{2}(\underline{\mathbf{y}}(1)+\overline{\mathbf{y}}(1))\Big) \geq 4\chi(1).
\end{equation}

With the network dynamics given by $f_\ast$, we therefore obtain
\begin{align}
  \chi(2) &= {\rm dist}(\mathbf{x}_{\theta_2}([\overline{s}]_2), \mathcal{I}_1)\nonumber\\
  &\geq {\rm dist}(\mathbf{x}_{d_1}(2), \mathcal{I}_1)\nonumber\\
  &\geq
  {\rm dist}(\mathbf{x}_{d_1}(2),(\underline{\mathbf{y}}(1)+\overline{\mathbf{y}}(1))/2)-|\mathcal{I}_1|/2\nonumber\\
  &\geq {\rm dist}\Big(\sum\limits_{j\in \mathrm{N}_{d_1}} a_{d_1j} f_\ast(\mathbf{x}_j(1)) + \mathbf{u}_{d_1}(1) + \mathbf{d}_{d_1}(1), (\underline{\mathbf{y}}(1)+\overline{\mathbf{y}}(1))/2\Big)-|\mathcal{I}_1|/2\nonumber\\
  &\geq 4\chi(1)-(|\mathcal{I}_0|+| \mathcal{R}_1| +| \mathcal{L}_1|)/2\nonumber\\
  &\geq  4\chi(1) - |\mathcal{I}_0|/2 - \chi(1) \nonumber\\
  &=  3\chi(1) - |\mathcal{I}_0|/2,\label{eqn:R2}
\end{align}
which is greater than $0$ indicating $\mathbf{x}_{d_1}(2)\notin \mathcal{I}_1$.

{\it\noindent Step 3:} Finally,  we proceed the analysis recursively and then obtain
\begin{align}
  \chi(t+1) &= {\rm dist}\big(\mathbf{x}_{\theta_{t+1}}([\overline{s}]_{t+1}), \mathcal{I}_t \big)\nonumber\\
  &\geq {\rm dist}\big(\mathbf{x}_{d_t}(t+1), \mathcal{I}_t\big)\nonumber\\
  &\geq
  {\rm dist}\big(\mathbf{x}_{d_t}(t+1),(\underline{\mathbf{y}}(t)+\overline{\mathbf{y}}(t))/2\big)-|\mathcal{I}_t|/2\nonumber\\
  &\geq {\rm dist}\Big(\sum\limits_{j\in \mathrm{N}_{d_t}} a_{d_tj} f_\ast(\mathbf{x}_j(t)) + \mathbf{u}_{d_t}(t) + \mathbf{d}_{d_t}(t), (\underline{\mathbf{y}}(t)+\overline{\mathbf{y}}(t))/2\Big)-|\mathcal{I}_t|/2\nonumber\\
  &\geq 4\chi(t)-\Big(|\mathcal{I}_0|+\sum\limits_{s=1}^t\big(| \mathcal{R}_s| +| \mathcal{L}_s| \big) \Big) \nonumber \\
  &\geq  4\chi(t) - |\mathcal{I}_0|/2 -\sum\limits_{s=1}^t \chi(s). \label{eqn:Rt}
\end{align}

Denote $E_0 = |\mathcal{I}_0|/2$ and $E_t=|\mathcal{I}_0|/2+\sum\limits_{s=1}^t \chi(s)$ for $t\geq 1$.  Then (\ref{eqn:R2}) (\ref{eqn:Rt}) can be written as
\begin{equation*}
  E_{t+1} - E_t \geq 4(E_{t} - E_{t-1}) -E_{t}, \ t\geq 1
\end{equation*}
or equivalently,
\begin{equation*}
  E_{t+1} - 2E_t\geq 2(E_t-2E_{t-1}),\  t\geq 1.
\end{equation*}
Therefore,
\begin{equation*}
  E_{t+1} - 2E_t\geq 2^t(E_1-2E_0)=2^t(\chi(1)-|\mathcal{I}_0|/2)>0.
\end{equation*}
This implies $
  E_{t+1}> 2E_t$, which in turn leads to $\chi(t)>0$ for all $t\geq 0$ and $  \sum\limits_{s=1}^t \chi(s) \rightarrow +\infty$ as time tends to infinity. The network dynamics therefore must diverge and we have concluded the proof of Theorem \ref{thm:impossibility}.

\subsection{Proof of Theorem \ref{thm:LIDD}}

For any $t\geq 0$, we define $ \overline{\mathbf{x}}_i(t) : = \max\limits_{0\leq s \leq t}\mathbf{x}_i(s)$, $\underline{\mathbf{x}}_i(t) : = \min\limits_{0\leq s \leq t}\mathbf{x}_j(s)$, and further
$\mathcal{I}_t^i := [\underline{\mathbf{x}}_i(t), \overline{\mathbf{x}}_i(t)]$
as the minimal interval containing all node states of node $i$ up to time $t$. We define $ \mathcal{R}_t^{i}:=(\overline{\mathbf{x}}_i(t-1), \overline{\mathbf{x}}_i(t)]$ and
$ \mathcal{L}_t^{i}:=[\underline{\mathbf{x}}_i(t) , \underline{\mathbf{x}}_i(t-1))$.
It is easy to observe that
\begin{equation}\label{eqn:decomIi}
\mathcal{I}_t^i = \{\mathbf{x}_i(0)\} \mcup(\bigcup\limits_{s=1}^{t}  \mathcal{R}_s^{i})\mcup (\bigcup\limits_{s=1}^{t}  \mathcal{L}_t^{i}),
\end{equation}
and $ \mathcal{R}_s^{i}$, $ \mathcal{L}_s^{i}$, $s = 1,\dots, t$, are disjoint.
Thus,
\begin{equation}\label{eqn:sumIi}
  |\mathcal{I}_t^i | =\sum\limits_{s=1}^{t}( | \mathcal{R}_s^{i}| + | \mathcal{L}_s^{i}|).
\end{equation}

The proof is divided into a few steps.

{\it\noindent Step 1:} In this step, we give an estimation of the difference between $\mathbf{x}_j(t)$ and $\mathbf{x}_{[\mathrm{v}_{ij}]_t}([s_{ij}]_t)$.

If $\mathbf{x}_j(t)\notin \mathcal{I}^j_{t-1}$, $t\geq 1$,
\begin{equation*}
|\mathbf{x}_j(t)-\mathbf{x}_{[\mathrm{v}_{ij}]_t}([s_{ij}]_t)| \leq \max\{| \mathcal{R}_t^{j}|,| \mathcal{L}_t^{j}|\}.
\end{equation*}
If $\mathbf{x}_j(t)\in \mathcal{I}^j_{t-1}$, $t\geq 1$, by (\ref{eqn:decomIi}), $\mathbf{x}_j(t)$ is contained in some $ \mathcal{R}_s^{j}$, $ \mathcal{L}_s^{j}$ or equal to $\mathbf{x}_i(0)$. Then, we have
\begin{equation*}
|\mathbf{x}_j(t)-\mathbf{x}_{[\mathrm{v}_{ij}]_t}([s_{ij}]_t)| \leq \max\{| \mathcal{R}_s^{j}|,| \mathcal{L}_s^{j}|\} \text{ for some }s\leq t-1.
\end{equation*}
Therefore, for any $i\in \mathrm{V}$, $j\in\mathrm{N}_i$ and $t\geq 1$,
\begin{equation}\label{eqn:differencei}
|\mathbf{x}_j(t)-\mathbf{x}_{[\mathrm{v}_{ij}]_t}([s_{ij}]_t)|\leq \max\limits_{1\leq s\leq t}\{| \mathcal{R}_s^{j}|,| \mathcal{L}_s^{j}|\}.
\end{equation}

{\it\noindent Step 2:} In this step, we find a recursive estimation of $ \mathcal{R}_{t+1}^{i}$ and $ \mathcal{L}_{t+1}^{i}$. We note for each $i\in\mathrm{V}$ that
\begin{equation*}
\left\{\begin{array}{lll}
| \mathcal{R}_{t+1}^{i}| &= 0, & \text{ if } \mathbf{x}_i(t+1) \leq \overline{\mathbf{x}}_i(t),\\
| \mathcal{R}_{t+1}^{i}| &= \mathbf{x}_i(t+1)-\overline{\mathbf{x}}_i(t),   & \text{ if } \mathbf{x}_i(t+1) > \overline{\mathbf{x}}_i(t).
\end{array}
\right.
\end{equation*}
We can thus conclude
\begin{align}
| \mathcal{R}_{t+1}^{i}|
&=  \big(\mathbf{x}_i(t+1)-\overline{\mathbf{x}}_i(t) \big)^+ \nonumber\\
&= \Big(\mathbf{x}_i(t+1)-\mathbf{x}_i(0)-\sum\limits_{s=1}^t| \mathcal{R}_s^{i}|\Big)^+ \nonumber\\
&\leq  \Big(|\mathbf{x}_i(t+1)-\mathbf{x}_i(0)|-\sum\limits_{s=1}^t| \mathcal{R}_s^{i}|\Big)^+.\label{eqn:deltaIi+}
\end{align}
According to (\ref{eqn:evol}) and (\ref{eqn:distributecontrol}),
\begin{align}
&|\mathbf{x}_i(t+1)-\mathbf{x}_i(0)| \nonumber\\
& = \Big|\sum_{j\in\mathrm{N}_i} a_{ij} f(\mathbf{x}_j(t)) + \mathbf{u}_i(t) +\mathbf{d}_i(t) - \mathbf{x}_i(0) \Big|\nonumber \\
& =   \Big |\sum\limits_{j\in\mathrm{N}_i} a_{ij} f(\mathbf{x}_j(t))-\sum\limits_{j\in\mathrm{N}_i} a_{ij} \mathbf{z}_{[\mathrm{v}_{ij}]_t}([s_{ij}]_t+1) +\mathbf{d}_i(t) \Big|\nonumber\\
& \leq  \sum\limits_{j\in\mathrm{N}_i} |a_{ij}|\Big(|f(\mathbf{x}_j(t))-f(\mathbf{x}_{[\mathrm{v}_{ij}]_t}([s_{ij}]_t)| + |f(\mathbf{x}_{[\mathrm{v}_{ij}]_t}([s_{ij}]_t)-\mathbf{z}_{[\mathrm{v}_{ij}]_t}([s_{ij}]_t+1)|\Big) +|\mathbf{d}_i(t)| \nonumber \\
& \leq  \sum\limits_{j\in\mathrm{N}_i} |a_{ij}|\Big(|f(\mathbf{x}_j(t))-f(\mathbf{x}_{[\mathrm{v}_{ij}]_t}([s_{ij}]_t)|+e_\ast\Big) +e_\ast \|A_{\mathrm{G}}\|_{\infty} +w_\ast
\nonumber\\
 &\leq  \sum\limits_{j\in\mathrm{N}_i} |a_{ij}|\big((L+\eta)\max\limits_{1\leq s\leq t}\{| \mathcal{R}_s^{j}|,| \mathcal{L}_s^{j}|\}+c+e_\ast\big) +e_\ast \|A_{\mathrm{G}}\|_{\infty}+w_\ast \nonumber\\
& \leq  (L+\eta)\sum\limits_{j\in\mathrm{N}_i}\big( |a_{ij}|\max\limits_{1\leq s\leq t}\{| \mathcal{R}_s^{j}|,| \mathcal{L}_s^{j}|\}\big) +w_i,\label{eqn:xkt+1final}
\end{align}
where $w_i= (c+2e_\ast) \|A_{\mathrm{G}}\|_{\infty} +w_\ast$. The third inequality is derived from (\ref{eqn:differencei}) and Lemma \ref{lemma-function}.
Therefore, by (\ref{eqn:deltaIi+}) and (\ref{eqn:xkt+1final}) and  choosing $\omega=\max\limits_{i\in \mathrm{V}}\omega_i$, we have
\begin{align}
| \mathcal{R}_{t+1}^{i}|  &\leq  \Big((L+\eta)\sum\limits_{j\in\mathrm{N}_i}\big( |a_{ij}|\max\limits_{1\leq s\leq t}\{| \mathcal{R}_s^{j}|,| \mathcal{L}_s^{j}|\}\big) +w-\sum\limits_{s=1}^t| \mathcal{R}_s^{i}|\Big) ^+.
\label{eqn:finalBi+}
\end{align}
Using the same method, we also have,
\begin{align}
| \mathcal{L}_{t+1}^{i}|  &\leq  \Big((L+\eta)\sum\limits_{j\in\mathrm{N}_i}\big( |a_{ij}|\max\limits_{1\leq s\leq t}\{| \mathcal{R}_s^{j}|,| \mathcal{L}_s^{j}|\}\big) +w-\sum\limits_{s=1}^t| \mathcal{L}_s^{i}|\Big) ^+.
\label{eqn:finalBi-}
\end{align}

{\it\noindent Step 3:} This step provides the final piece of the proof. By the definition of $\|A_{\mathrm{G}}\|_{\dag}$, we know that $\sum_{t=1}^\infty \big(\mathcal{R}_{t+1}^{i}+\mathcal{L}_{t+1}^{i}\big)<\infty$
for all $i\in\mathrm{V}$. This proves stabilization of the network dynamics and concludes the proof of the desired theorem.

\subsection{Proof of Theorem \ref{thm:max}}
Recall that $\mathcal{I}_t:= [\underline{\mathbf{y}}(t), \overline{\mathbf{y}}(t)]$, $|\mathcal{I}_t|=\overline{\mathbf{y}}(t)-\underline{\mathbf{y}}(t)$,  $ \mathcal{R}_t=(\overline{\mathbf{y}}(t-1), \overline{\mathbf{y}}(t)]$, and $ \mathcal{L}_t=[\underline{\mathbf{y}}(t) , \underline{\mathbf{y}}(t-1))$. The Eq. (\ref{eqn:decomB}) holds. By their definition there holds
\begin{align*}
  \overline{\mathbf{y}}(t) = \max\limits_{0\leq s\leq t}\overline{\mathbf{x}}(s), \quad  \underline{\mathbf{y}}(t) = \min\limits_{0\leq s\leq t}\underline{\mathbf{x}}(s).
\end{align*}
Because all nodes know $\overline{\mathbf{x}}(t)$  and $\underline{\mathbf{x}}(t)$ before time $t+1$, they know $\overline{\mathbf{y}}(t)$  and $\underline{\mathbf{y}}(t)$ too.

For  $j\in \mathrm{N}_i$, if $\mathbf{x}_j(t)\notin \mathcal{I}_{t-1}$, $t\geq 1$, we have
\begin{equation*}
\Big|\mathbf{x}_j(t)-{\arg\min}_{x\in\mathcal{X}_i(t)}\big\{|\mathbf{x}_j(t) -x |\big\}\Big| \leq \max\{| \mathcal{R}_t|,| \mathcal{L}_t|\}.
\end{equation*}
If $\mathbf{x}_j(t)\in \mathcal{I}_{t-1}$, $t\geq 1$, by (\ref{eqn:decomB}), $\mathbf{x}_j(t)$ is contained in some $ \mathcal{R}_s$, $ \mathcal{L}_s$ or $\mathcal{I}_0$. Thus,
\begin{equation*}
\Big|\mathbf{x}_j(t)-{\arg\min}_{x\in\mathcal{X}_i(t)}\big\{|\mathbf{x}_j(t) -x |\big\}\Big| \leq \max\{| \mathcal{R}_s|,| \mathcal{L}_s|, |\mathcal{I}_0|\} \text{ for some }s\leq t-1.
\end{equation*}
Therefore, for $t\geq 1$,
\begin{equation}\label{eqn:differencesemi}
\Big|\mathbf{x}_i(t)-{\arg\min}_{x\in\mathcal{X}_i(t)}\big\{|\mathbf{x}_j(t) -x |\big\}\Big|\leq \max\big\{\max\limits_{1\leq s\leq t}\{| \mathcal{R}_t|,| \mathcal{L}_t|\}, |\mathcal{I}_0|\big\}.
\end{equation}

Note that (\ref{eqn:deltaB+}) is irrelevant to the controller, so they also hold in this case.
According to (\ref{eqn:evol}) and (\ref{eqn:maxenhancedcontrol}),
\begin{align*}
&\Big|\mathbf{x}_i(t+1)-\frac{1}{2}(\underline{\mathbf{y}}(t)+\overline{\mathbf{y}}(t) )\Big|\nonumber\\
 & = \Big|\sum_{j\in\mathrm{N}_i} a_{ij} f(\mathbf{x}_j(t)) + \mathbf{u}_i(t) +\mathbf{d}_i(t) -\frac{1}{2}(\underline{\mathbf{y}}(t)+\overline{\mathbf{y}}(t) ) \Big|\nonumber \\
& = \Big  |\sum\limits_{j\in\mathrm{N}_i} a_{ij} f(\mathbf{x}_j(t))-\sum\limits_{j\in\mathrm{N}_i} a_{ij} \widehat{f}(\mathbf{x}_j(t))  +\mathbf{d}_i(t)\Big|\nonumber\\
& \leq  \sum\limits_{j\in\mathrm{N}_i} |a_{ij}|\Big(\Big|f(\mathbf{x}_j(t))-f\Big({\arg\min}_{x\in\mathcal{X}_i(t)}\big\{|\mathbf{x}_j(t) -x |\big\}\Big)\Big| \nonumber\\& \quad + \Big|f\Big({\arg\min}_{x\in\mathcal{X}_i(t)}\big\{|\mathbf{x}_j(t) -x |\big\}\Big)-\mathcal{K}_i^t\Big({\arg\min}_{x\in\mathcal{X}_i(t)}\big\{|\mathbf{x}_j(t) -x |\big\} \Big)\Big|\Big) +e_\ast\|A_{\mathrm{G}}\|_{\infty} +w_\ast \nonumber \\
& \leq  \sum\limits_{j\in\mathrm{N}_i} |a_{ij}|\Big|f(\mathbf{x}_j(t))-f\Big({\arg\min}_{x\in\mathcal{X}_i(t)}\big\{|\mathbf{x}_j(t) -x |\big\}\Big)\Big|+2e_\ast\|A_{\mathrm{G}}\|_{\infty} +w_\ast.\nonumber\\
&\leq  \|A_{\mathrm{G}}\|_{\infty}(L+\eta)\max\{\max\limits_{1\leq s\leq t}| \mathcal{R}_s|,\max\limits_{1\leq s\leq t}| \mathcal{L}_s|, |\mathcal{I}_0|\}+E_\ast,
\end{align*}
where $E_\ast= (c+2e_\ast)\|A_{\mathrm{G}}\|_{\infty} +w_\ast$. This is exactly (\ref{eqn:vcenter}). Therefore, Eq. (\ref{eqn:finalB+}) continues to hold. From here and by invoking Lemma \ref{lemma:finitesum}, the desired theorem can be established.

\section{Conclusions}\label{sec:conclusions}
This paper proposes a framework for studying  the fundamental limitations of feedback mechanism in dealing with uncertainties over network systems. The study of maximum capability of feedback control was pioneered in Xie and Guo (2000) for simple scalar system with discrete-time dynamics. We have successfully extended such effort  to a network setting, where nodes with unknown and nonlinear   dynamics hold interconnections  through a directed interaction graph.   Using information structure and decision pattern as  criteria, three classes   of feedback laws over such networks were defined, under which  critical or sufficient feedback capacities were established, respectively. These preliminary  results reveal a promising path  towards clear descriptions of feedback capabilities  over complex network systems, many important problems  yet remain open.

First of all, the fundamental limitations established in the current  work are for generic graphs. How a given  structure influences feedback capacity over networks has not been answered and it is  a challenging question. Next, the model under investigation assumes measurement and control at  all nodes, therefore it is very  interesting to ask the same feedback capacity questions  when only a subset of nodes can be monitors of the information flow and another subset of nodes can be controlled as anchors \cite{Magnus-SIAM-2009, Olshevsky-TCNS-2014}.  Finally, parametric network model as generalizations to the work of  \cite{Li-Guo-Automatica-2010} and \cite{Sokolov-Automatica-2016} would be intriguing because such a model will certainly  yield a strong connection between distributed estimation and distributed control.

\section*{Appendix. Positivity of  $\|A_{\mathrm{G}}\|_{\dag}$}

The positivity of  $\|A_{\mathrm{G}}\|_{\dag}$ for nontrivial graphs is implied by the following lemma.
\begin{lemma}\label{lemma:positivity}
$\|A_{\mathrm{G}}\|_{\dag}\geq 1/\|A_{\mathrm{G}}\|_{\infty}$.
\end{lemma}
{\noindent \em Proof:}  Introduce node set $
\mathrm{V}^{\infty}_p = \big\{i\in\mathrm{V}: \sum_{s=1}^\infty p^i_s=\infty\big\}$ and  $
\mathrm{V}^{\infty}_q = \big\{i\in\mathrm{V}: \sum_{s=1}^\infty q^i_s=\infty\big\}$.
Assume that $\mathrm{V}^{\infty}_p$ is nonempty. Let us suppose for the moment  $$
 {B}^\ast:=\max\big\{p_s^j, q_s^j: 1\leq s < \infty, j\in\mathrm{V} \big\}
$$
is a finite number. Then there exists $t_1>0$ such that
$\sum_{s=1}^t p^i_s>\omega+ {B}^\ast$ for all $t>t_1$, $i\in\mathrm{V}^{\infty}_p$. Consequently, letting  $M=1/\|A_{\mathrm{G}}\|_{\infty}$ in  (\ref{inequality}) we know for all $i\in\mathrm{V}^{\infty}_p$ and  $t>t_1$ that
      \[
      M\sum\limits_{j\in\mathrm{N}_i} |a_{ij}|\max\limits_{1\leq s\leq t}\{p_s^{j},q_s^{j}\} +\omega-\sum\limits_{s=1}^t p^i_{s} \leq M\|A_{\mathrm{G}}\|_{\infty} {B}^\ast+\omega-(\omega+ {B}^\ast)=0.
      \]
      Therefore, $p^i_{t+1}=0$ for $i\in\mathrm{V}^{\infty}_p$ and all $t>t_1$, implying  $\sum_{s=1}^\infty p^i_s<\infty$. This is not possible and therefore  $ {B}^\ast=\infty$.

Note that, from the definition of $\mathrm{V}^{\infty}_p$ and $\mathrm{V}^{\infty}_q$, there must  exist $C^*>\omega$ such that $
 \sum_{s=1}^\infty p^i_s< {C}^\ast$
for all $i\in \mathrm{V}\backslash\mathrm{V}^{\infty}_p$, and $
 \sum_{s=1}^\infty q^i_s< {C}^\ast$
for all $i\in \mathrm{V}\backslash\mathrm{V}^{\infty}_q$. %These imply that for any $s>0$
%\begin{equation}\label{eqn:psC}
%  p_s^i< {C}^\ast, \quad \mbox{$i\in\backslash\mathrm{V}^{\infty}_p$},
%\end{equation}
%and
%\begin{equation}\label{eqn:qsC}
%  q_s^i< {C}^\ast, \quad \mbox{$i\in\backslash\mathrm{V}^{\infty}_q$}.
%\end{equation}
%
%
%
%
%Further,
We can also find $t_2>0$ such that when $t>t_2$,
$\sum_{s=1}^t p^i_s> {C}^\ast$ for all $i\in\mathrm{V}^{\infty}_p$ and
$\sum_{s=1}^t q^i_s> {C}^\ast$
for all $i\in\mathrm{V}^{\infty}_q$. We select  $t_3=t_2+1$ and define $$
 {D}^\ast: =\max\big\{p_s^j, q_s^j: 1\leq s \leq  t_3, j\in\mathrm{V}\big\}
$$
which should be no less than $C^*$. Then, for $i\in\mathrm{V}^{\infty}_p$ there holds that
\begin{equation}\label{eqn:ip}
      M\sum\limits_{j\in\mathrm{N}_i} |a_{ij}|\max\limits_{1\leq s\leq t_3}\{p_s^{j},q_s^{j}\} +\omega-\sum\limits_{s=1}^{t_3} p^i_{s}\leq M\|A_{\mathrm{G}}\|_{\infty} {D}^\ast +\omega -  {C}^\ast \leq  {D}^\ast,
\end{equation}
and for $i\in\mathrm{V}^{\infty}_q$ there holds that
\begin{equation}\label{eqn:iq}
M\sum\limits_{j\in\mathrm{N}_i} |a_{ij}|\max\limits_{1\leq s\leq t_3}\{p_s^{j},q_s^{j}\} +\omega-\sum\limits_{s=1}^{t_3} q^i_{s}\leq M\|A_{\mathrm{G}}\|_{\infty} {D}^\ast +\omega -  {C}^\ast \leq  {D}^\ast.
\end{equation}
Thus,
$p^i_{t_3+1}\leq  {D}^\ast$ for $i\in\mathrm{V}^{\infty}_p$ and $q^i_{t_3+1}\leq  {D}^\ast$ for $i\in\mathrm{V}^{\infty}_q$.

This leaves  $$
 {D}^\ast =\max\big\{p_s^j, q_s^j: 1\leq s \leq  t_3+1, j\in\mathrm{V}\big\}
 $$
being  the only possibility.  The recursive inequalities further ensure $ {D}^\ast= \max \big\{p_s^j, q_s^j: 1\leq s < \infty, j\in\mathrm{V} \big\}<\infty$, contradicting   $ {B}^\ast=\infty$. As a result, $\mathrm{V}^{\infty}_p$ is an empty set. For the same reason $\mathrm{V}^{\infty}_q$ is an empty set.  In other words,  if $M=1/\|A_{\mathrm{G}}\|_{\infty}$, then for any $\omega>0$ Eq. (\ref{inequality}) implies $\sum\limits_{t=1}^\infty (p^i_t+q^i_t)<\infty$  for all  $i\in \mathrm{V}$. We have now proved  $\|A_{\mathrm{G}}\|_{\dag}\geq 1/\|A_{\mathrm{G}}\|_{\infty}$ by the definition of $\|A_{\mathrm{G}}\|_{\dag}$. \hfill$\square$


\begin{thebibliography}{10}
%\bibitem{lai1987}
%T.~L. Lai.
%\newblock Adaptive treatment allocation and the multi-armed bandit problem.
%\newblock {\em The Annals of Statistics}, 15(3):1091--1114, 09 1987.
%
%\bibitem{lai1985}
%T. L. Lai and H.~Robbins.
%\newblock Asymptotically efficient adaptive allocation rules.
%\newblock {\em Advances in Applied Mathematics}, 6(1):4--2, 1985.

\bibitem{Maxwell-1968}J. C. Maxwell, ``On governors," {\em Proc. R. Soc. Lond.}, vol. 16, pp. 270--283, 1868.


\bibitem{Franklin-Book-2015}G. F. Franklin, J. D. Powell, and A. Emami-Naeini. {\em Feedback Control of Dynamic Systems}. 7th Edition, Prentice Hall, 2015.



\bibitem{Basar-Book-1991} T. Basar and P. Bernhard. {\em $H^\infty$ Optimal Control and Related Minimax
Design Problems: A Dynamic Game Approach}. Boston, MA:
Birkhauser, 1991.





\bibitem{Zames-1966} G. Zames, ``On the input–output stability of time-varying nonlinear systems Part I: Conditions using concepts of loop gain, conicity, and positivity," {\em  IEEE Trans. Automat. Contr.}, vol. 11, pp. 228--238, 1966.


\bibitem{Doyle-1982} J. C. Doyle,``Analysis of feedback systems with structured uncertainties,"
{\em IEE Proc.}, vol. 133, pp. 45--56, 1982.


\bibitem{Khammash-1991} M. Khammash and J. B. Pearson. ``Performance robustness of discrete-time
systems with structured uncertainty," {\em IEEE Trans.  Autom. Contr.},
vol. 36, 398--412, 1991.

\bibitem{Zhou-Book-1996} K. Zhou, J. C. Doyle, and K. Glover. {\em  Robust and Optimal Control}. Prentice Hall, 1996.



\bibitem{Matt-Book-1999} J. W. Helton and M. R. James. {\em Extending $H^\infty$ Control to Nonlinear
Systems.} Frontiers in Applied Mathematics, SIAM, 1999.


%%%%%%%%%%%%%%%%%%%%%%%%%%%%%%%%%%%%%%%%%%%%%%%%%%%%%%%%%%%%%%% Adaptive Control

\bibitem{Astrom-Automatica-1973} K. J. {\AA}str\"{o}m and B. Wittenmark, ``On self tuning regulators," vol. 9, no. 2, pp. 185--199, 1973.
\bibitem{Mareels-Automatica-1992} I. M. Y. Mareels, H. B. Penfold, and R. J. Evans, ``Controlling nonlinear
time-varying systems via Euler approximations," {\em Automatica}, vol. 28,
pp. 681--696, 1992.

\bibitem{Mareels-Automatica-1998}  E. Skafidas, A. Fradkov, R. J. Evans, and I. M. Y. Mareels, ``Trajectory-approximation-based adaptive control for nonlinear systems under
matching conditions," {\em Automatica}, vol. 34, pp. 287--299, 1998.

\bibitem{Byrnes} C. L. Byrnes and W. Lin, ``Losslessness, feedback equivalence and the
global stabilization of discrete-time nonlinear systems," {\em IEEE Trans.
Automat. Contr.}, vol. 39, pp. 83--97, 1994.


\bibitem{Astrom-Book-1995} K. J. {\AA}str\"{o}m and B. Wittenmark. {\em Adaptive Control}, 2nd Edition. Reading,
MA: Addison-Wesley, 1995.

\bibitem{Krstic-Book-1995}M. Krsti\'{c}, I. Kanellakopoulos, and P. V. Kokotovi\'{c}. {\em Nonlinear and Adaptive
Control Design}. New York: Wiley, 1995.



%%%%%%%%%%%%%%%%%%%%%%%%%%%%%%%%%%%%%%%%%%%%%%%%%%%%%%%%%%%%%%% Robust Control




%%%%%%%%%%%%%%%%%%%%%%%%%%%%%%%%%%%%%%%%%%%%%%%%%%%%%%%%%%%%%%% Control of Network Systems
\bibitem{Wong-TAC-1999}W. S. Wong and R. W. Brockett, ``Systems with finite communication
bandwidth constraints: II. Stabilization with
limited information feedback," {\em  IEEE Trans.
Autom. Contr.}, vol. 44, no. 5, pp. 1049--1053, 1999.


\bibitem{Nair-SIAM-2004} G. N. Nair and R. J. Evans, ``Stabilizability of stochastic linear systems with finite feedback data rates," {\em SIAM Journal on Control and Optimization}, vol. 43, no. 2, pp. 413--436, 2004.


  \bibitem{Liberzon-TAC-2005}  D. Liberzon and J. P. Hespanha, ``Stabilization
of nonlinear systems with limited information
feedback," {\em IEEE Trans. Autom. Contr.}, vol. 50,
no. 6, pp. 910--915, 2005.

\bibitem{Nair-PIEEE-2007} G. N. Nair, F. Fagnani, S. Zampieri, R. J. Evans, ``Feedback control under data rate constraints: an overview," {\em Proceedings of the IEEE},  Special issue on technology of networked control systems, vol. 95, no. 1, pp. 108--137,  2007.


\bibitem{Nair-TAC-2013} G. N. Nair, ``A nonstochastic information theory for communication and state estimation," {\em IEEE Trans.  Autom. Contr.},  vol. 58, no. 6, pp. 1497--1510, 2013.



    \bibitem{Schenato-PIEEE-2007} L. Schenato,  B. Sinopoli,  M. Franceschetti,  K. Poolla, and
S. S. Sastry, ``Foundations of control and
estimation over lossy networks," {\em Procedings of IEEE},  vol. 95, no. 1, pp. 163--187,  2007.

    \bibitem{Dahleh-TAC-2008} N. C. Martins  and M. A. Dahleh, ``Feedback control in the presence of noisy
channels: "Bode-like fundamental
limitations of performance," {\em IEEE Trans. Autom. Contr.},  vol. 53, no 7, pp. 1604--1615, 2008.




\bibitem{Jadbabaie-TAC-2003} A. Jadbabaie, J. Lin, and A. S. Morse,  ``Coordination of groups of mobile
autonomous agents using nearest neighbor rules," {\em IEEE Trans. Autom.
Control}, vol. 48, no. 6, pp. 988--1001, 2003.

\bibitem{saber2004} R. Olfati-Saber and R. M. Murray, ``Consensus problems in networks of
agents with switching topology and time-delays," {\em IEEE Trans.  Autom. Contr.}, vol. 49, no. 9, pp. 1520--1533,  2004.


\bibitem{Bullo-TAC-2007} S. Martinez, J. Cortes, and F. Bullo, ``Motion coordination with distributed information," {\em  IEEE Control Syst. Mag.}, vol. 27, no. 4, pp. 75--88,
2007.

\bibitem{Nedic-TAC-2010} A. Nedic, A. Ozdaglar and P. A. Parrilo, ``Constrained consensus and
optimization in multi-agent networks," {\em IEEE Trans. Autom. Control}, vol.
55, no. 4, pp. 922--938, 2010.

\bibitem{Mou-TAC-2015} S. Mou, J, Liu, and A. S. Morse, ``A distributed algorithm for solving
a linear algebraic equation," {\em IEEE Trans. Autom. Control}, vol. 60, no. 11,  pp. 2863--2878, 2015.

\bibitem{Shi-TAC-LAE} G. Shi, B. D. O. Anderson, and U. Helmke, ``Network flows that solve linear equations," {\em IEEE Trans.  Autom. Contr.}, vol. 62, no. 6, pp. 2659--2674, 2017.



\bibitem{Magnus-Book-2010} M. Mesbahi and M. Egerstedt. {\em Graph Theoretic Methods in Multiagent
Networks}. Princeton University Press, 2010.





\bibitem{Shi-TAC-Quantum} G. Shi, D. Dong, I. R. Petersen, and K. H. Johansson, ``Reaching a quantum consensus: master equations that generate symmetrization and synchronization," {\em IEEE Trans.  Autom. Contr.}, vol. 61, no. 2, pp. 374--387,  2016.


\bibitem{Magnus-SIAM-2009} A. Rahmani, M. Ji, M. Mesbahi, and M. Egerstedt, ``Controllability of multi-agent systems from a graph-theoretic
perspective," {\em  SIAM J. Control Optim.}, vol. 48, no. 1, pp. 162--186, 2009.




\bibitem{Barabasi-Nature-2011} Y.-Y. Liu, J.-J. Slotine, and A.-L. Barabasi, ``Controllability of complex networks," {\em Nature}, 473:167--173, 2011.


\bibitem{Olshevsky-TCNS-2014} A. Olshevsky, ``Minimal controllability problems," {\em  IEEE Trans. Control of Network Systems}, vol. 1, no. 3, pp. 249--258,
2014.


%%%%%%%%%%%%%%%%%%%%%%%%%%%%%%%%%%%%%%%%%%%%%%%%%%%%%%%%%%%%%%% Feedback Capacity
\bibitem{xie-guo2000}
L.-L. Xie and L. Guo, ``How much uncertainty can be Dealt with by
feedback?" {\em IEEE Trans.  Autom. Contr.}, vol. 45, no. 12, pp. 2203--2217, 2000.



\bibitem{Li-Guo-Automatica-2010} C. Y. Li and L. Guo,  ``A new critical theorem for adaptive nonlinear stabilization,"
{\em Automatica}, vol. 46, pp. 999--1007, 2010.

\bibitem{Li-Guo-TAC-2011} C. Y. Li and L. Guo,``On feedback capability in nonlinearly parameterized
uncertain dynamical systems," {\em IEEE Trans.  Autom. Contr.}, vol. 56, pp.
2946--2951, 2011.

\bibitem{Huang-Guo-Automatica-2012} C. Huang and L. Guo, ``On feedback capability for a class of semiparametric
uncertain systems," {\em  Automatica}, vol. 48, pp. 873--878, 2012.

\bibitem{Sokolov-Automatica-2016}V. F. Sokolov, ``Adaptive stabilization of parameter-affine minimum-phase plants
under Lipschitz uncertainty," {\em Automatica}, vol. 73, pp. 64--70, 2016.




%%%%%%%%%%%%%%%%%%%%%%%%%%%%%%%%%%%%%%%%%%%%%%%%%%%%%%%%%%%%%%%%% Max Consensus

\bibitem{jadbabaie2006} A. Tahbaz-Salehi and A. Jadbabaie, ``A one-parameter family of distributed
consensus algorithms with boundary: From shortest paths to
mean hitting times," in {\em 45th IEEE Conference on Decision and Control}, pp. 4664--4669, 2006.
%
%\bibitem{cortes} J. Cort\'{e}s, ``Distributed algorithms for reaching consensus on general
%functions," {\em Automatica}, vol. 44, no. 3, pp. 401--420, 2008.
%

\bibitem{Giannini2013} S. Giannini, D. D. Paola, A. Petitti, and A. Rizzo, ``On the convergence
of the max-consensus protocol with asynchronous updates," in {\em 52nd
IEEE Conference on Decision and Control},   2013.


\bibitem{Iutzeler2012} F. Iutzeler, P. Ciblat, and J. Jakubowicz, ``Analysis of max-consensus
algorithms in wireless channels," {\em IEEE Trans. Signal Processing},
vol. 60, pp. 6103--6107,  2012.

\end{thebibliography}
\end{document}